\documentclass[conference]{IEEEtran}
\IEEEoverridecommandlockouts

\usepackage[utf8]{inputenc}
\usepackage[T1]{fontenc}
\usepackage{graphicx}
\usepackage{amssymb}
\usepackage{amsmath}
\usepackage{amsthm}
\usepackage{xspace}
\usepackage{mathtools}
\usepackage{xcolor}
\usepackage{url}
\usepackage[hidelinks]{hyperref}
\usepackage[nameinlink,capitalize,noabbrev]{cleveref}
\usepackage{bbold}
\usepackage{xfrac}
\usepackage{nicefrac}
\usepackage{microtype}
\usepackage{subfigure}
\usepackage{algorithm}
\usepackage[noend]{algorithmic}
\usepackage{booktabs}
\usepackage{multirow}
\usepackage{enumitem}
\usepackage{titletoc}
\usepackage[numbers]{natbib}

\makeatletter
 \newcommand{\newreptheorem}[2]{%
  \newtheorem*{rep@#1}{\rep@title}%
  \newenvironment{rep#1}[1]{%
   \def\rep@title{#2 \ref*{##1}}%
  \begin{rep@#1}}%
 {\end{rep@#1}}}
\makeatother

\newtheorem{theorem}{Theorem}
\newreptheorem{theorem}{Theorem}
\newtheorem{lemma}[theorem]{Lemma}
\newreptheorem{lemma}{Lemma}
\AddToHook{env/lemma/begin}{\crefalias{theorem}{lemma}}
\newtheorem{corollary}[theorem]{Corollary}
\newreptheorem{corollary}{Corollary}
\AddToHook{env/corollary/begin}{\crefalias{theorem}{corollary}}
\theoremstyle{definition}
\newtheorem{definition}[theorem]{Definition}
\AddToHook{env/definition/begin}{\crefalias{theorem}{definition}}
\theoremstyle{remark}
\newtheorem{example}[theorem]{Example}
\AddToHook{env/example/begin}{\crefalias{theorem}{example}}
\crefname{figure}{Fig.}{Figs.}
\crefname{subsubsubappendix}{Appendix}{Appendices}

\newcommand{\ppmlnameFct}[1]{#1{}SVM-SGD\xspace}
\newcommand{\softmaxnameFct}[1]{#1{}SoftmaxReg-SGD\xspace}
\newcommand{\ppmlname}{\ppmlnameFct{}}
\newcommand{\softmaxname}{\softmaxnameFct{}}
\newcommand{\softmaxslp}{SoftmaxReg\xspace}
\newcommand{\blindavg}{BlindAvg\xspace}

\usepackage{pgfplots}
\usepackage{tikzscale}
\pgfplotsset{compat=1.17}

\newcommand{\eps}{\varepsilon}
\newcommand{\set}[1]{\bgroup\left\{\,#1\,\right\}\egroup\xspace}
\newcommand{\myset}[1]{\ensuremath{\left\{#1\right\}}}
\newcommand{\noncolluding}{\ensuremath{t}\xspace}
\newcommand{\norm}[1]{\left\lVert#1\right\rVert}
\newcommand{\inner}[2]{\left\langle#1,#2\right\rangle}
\newcommand{\Abs}[1]{\left\lvert#1\right\rvert}
\DeclarePairedDelimiter\abs{\lvert}{\rvert}
\DeclareMathOperator*{\Eop}{\mathbb{E}}
\newcommand{\E}[1]{\operatorname{\mathbb{E}}\!\bgroup\left[#1\right]\egroup\xspace}
\newcommand{\avg}{\ensuremath{\mathrm{avg}}\xspace}
\DeclareMathOperator*{\argmin}{argmin}

\DeclareMathOperator*{\RV}{RV}
\DeclareMathOperator*{\pdf}{pdf}
\DeclareMathOperator*{\diag}{diag}
\DeclareMathOperator*{\erfc}{erfc}
\DeclareMathOperator*{\spn}{span}

\newcommand{\denseparagraph}[1]{\noindent\textbf{#1}\quad}

\begin{document}

\title{Private Blind Model Averaging -- \\ Distributed, Non-interactive, and Convergent
    \thanks{This work has been accepted for publication at the \emph{IEEE Conference on Secure and Trustworthy Machine Learning (SaTML)}. The final version will be available on IEEE Xplore.}
}

\author{
    \IEEEauthorblockN{1\textsuperscript{st} Moritz Kirschte}
    \IEEEauthorblockA{\textit{University of Luebeck}\\
    Lübeck, Germany \\
    m.kirschte@uni-luebeck.de}
    \and
    \IEEEauthorblockN{2\textsuperscript{nd} Sebastian Meiser}
    \IEEEauthorblockA{\textit{University of Luebeck}\\
    Lübeck, Germany \\
    sebastian.meiser@uni-luebeck.de}
    \and
    \IEEEauthorblockN{3\textsuperscript{rd} Saman Ardalan}
    \IEEEauthorblockA{\textit{UKSH Kiel}\\
    Kiel, Germany \\
    saman.ardalan@uksh.de}
    \and
    \IEEEauthorblockN{4\textsuperscript{th} Esfandiar Mohammadi}
    \IEEEauthorblockA{\textit{University of Luebeck}\\
    Lübeck, Germany \\
    esfandiar.mohammadi@uni-luebeck.de}
}

\maketitle

\begin{abstract}

Distributed differentially private learning techniques enable a large number of users to jointly learn a model without having to first centrally collect the training data. At the same time, neither the communication between the users nor the resulting model shall leak information about the training data. This kind of learning technique can be deployed to edge devices if it can be scaled up to a large number of users, particularly if the communication is reduced to a minimum: no interaction, i.e., each party only sends a single message. The best previously known methods are based on gradient averaging, which inherently requires many synchronization rounds.
A promising non-interactive alternative to gradient averaging relies on so-called output perturbation: each user first locally finishes training and then submits its model for secure averaging without further synchronization. We analyze this paradigm, which we coin blind model averaging (BlindAvg), in the setting of convex and smooth empirical risk minimization (ERM) like a support vector machine (SVM). While the required noise scale is asymptotically the same as in the centralized setting, it is not well understood how close BlindAvg comes to centralized learning, i.e., its utility cost.

We characterize and boost the privacy-utility tradeoff of BlindAvg with two contributions: 
First, we prove that BlindAvg converges towards the centralized setting for a sufficiently strong L2-regularization for a non-smooth SVM learner. Second, we introduce the novel differentially private convex and smooth ERM learner SoftmaxReg that has a better privacy-utility tradeoff than an SVM in a multi-class setting.
We evaluate our findings on three datasets (CIFAR-10, CIFAR-100, and Federated EMNIST) and provide an ablation in an artificially extreme non-IID scenario.

\end{abstract}


\section{Introduction}
    \begin{figure*}[!ht]
    \centerline{\includegraphics[width=0.97\textwidth]{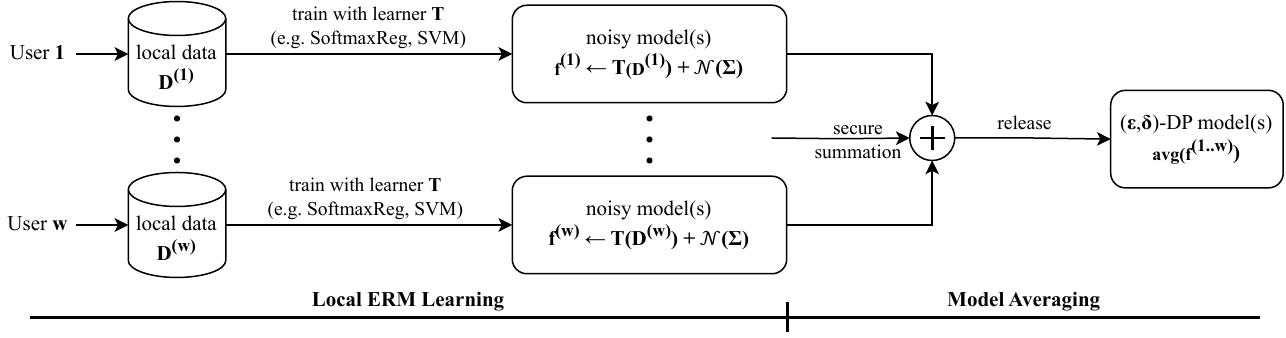}}
    \caption{\textbf{Schematic overview} of blind model averaging (\blindavg).}
    \label{fig:schematics}
\end{figure*}

Distributed privacy-preserving learning enables collaboratively training a machine learning model while satisfying strong privacy guarantees about the training data. Such learning techniques face two challenges: how to achieve comparable privacy-utility trade-offs to the centralized setting, and how to achieve scalability to a large number of users. One hurdle of scalability is the communication overhead which is optimal in the non-interactivity setting, i.e., each user only sends a single message without any further synchronization.

We study how a non-interactive approach, which we call \emph{blind model averaging} (\blindavg, cf. \Cref{fig:schematics}), satisfies these two challenges.
In \blindavg, we assume $w$ users, each with a local dataset $D^{(i)}$ of size $n = |D^{(i)}|$. First, each user locally trains a model $f^{(i)} = T_\xi(D^{(i)})$ using a convex and smooth empirical risk minimization (ERM) learner $T_\xi$, e.g., a support vector machine (SVM). After the training, each user locally perturbs the model parameters $f^{(i)}$ with Gaussian noise $\mathcal{N}$: $f_{\mathrm{priv}}^{(i)} = f^{(i)} + \mathcal{N}(0, \Sigma)$ (so-called output perturbation). Second, all perturbed models are securely aggregated (e.g. via a secure summation multi-party protocol \citep{bell2020secure}) and the result is released: $f_{\mathrm{priv}} = \frac{1}{w}\sum_{i=1}^w f_{\mathrm{priv}}^{(i)}$.
While it is known that \blindavg satisfies $\eps$-differential privacy---a state-of-the-art privacy notion---with guarantees just like in a centralized setting ($\eps \propto \Theta(\frac{1}{nw})$), the privacy-utility tradeoff of \blindavg is poorly understood.

The most prominent alternative to \blindavg is gradient averaging (GradAvg) \citep{abadi2016deep,McMahanMRHA17} which inherently requires synchronization rounds. In a simplified variant of GradAvg, all users noise and aggregate their locally computed gradients in each training iteration. Hence, GradAvg either involves a significant noise overhead and high communication overhead (`GradAvg' in \Cref{tab:realted_work_comp}) or cryptographic schemes without noise overhead but an even higher communication overhead (`GradAvg + SecSum' in \Cref{tab:realted_work_comp}). Due to these drawbacks, GradAvg and especially GradAvg+SecSum are unsuited for applications with end-user devices where users are unavailable for the whole training duration or have communication constraints, such as a metered internet connection.

\denseparagraph{Our Work.} 
We characterize the privacy-utility tradeoff of \blindavg in two problem statements, together with our respective contributions.
In \Cref{sec:intro_centralized}, we characterize for which problems \blindavg does not incur a significant utility cost for a large number of users.
In \Cref{sec:intro_softmaxreg}, we show for a well-known multi-class learner (Softmax Regression) that it satisfies the strict mathematical requirements of \blindavg.

\section{Problem Statements \& Contributions}
\subsection{Can \blindavg Approximate Centralized Learning?}\label{sec:intro_centralized}

\begin{figure*}[!ht]
    \centering
    \label{fig:intution_convergence_a}{\resizebox{0.325\textwidth}{!}{\input{plots/gen_convergance_intuition_SVM0.pgf}\unskip}}
    \label{fig:intution_convergence_b}{\resizebox{0.325\textwidth}{!}{\input{plots/gen_convergance_intuition_SVM1.pgf}\unskip}}
    \label{fig:intution_convergence_c}{\resizebox{0.325\textwidth}{!}{\input{plots/gen_convergance_intuition_3.pgf}\unskip}}
    \subfigure[SVM $f_1$ of User 1]{\label{fig:intution_convergence_e}{\resizebox{0.325\textwidth}{!}{\input{plots/gen_convergance_intuition2_SVM0.pgf}\unskip}}}
    \subfigure[SVM $f_2$ of User 2]{\label{fig:intution_convergence_f}{\resizebox{0.325\textwidth}{!}{\input{plots/gen_convergance_intuition2_SVM1.pgf}\unskip}}}
    \subfigure[Averaged / central SVM]{\label{fig:intution_convergence_g}{\resizebox{0.325\textwidth}{!}{\input{plots/gen_convergance_intuition2_3.pgf}\unskip}}}
    
    \caption{
        \textbf{Putting \blindavg to the extreme}: datasets \texttt{SynNonIID} (top) and \texttt{SynFail} (bottom).
        We plot: (a,b) the local 2d data of two classes (orange and blue) and SVMs (purple and red solid lines) of two users with their margins (dotted); (c) the averaged SVM (green) of these two users and a central SVM trained on the combined data (black).
    }
    \label{fig:intution_convergence}
\end{figure*}

For understanding the utility-privacy tradeoffs of \blindavg, we analyze in \Cref{fig:intution_convergence} two extreme scenarios (without noise): a strongly-biased non-IID scenario where \blindavg matches (\texttt{SynNonIID}) and where it fails to match the centralized accuracy (\texttt{SynFail}). \texttt{SynFail} represents a class of problems where \blindavg has its limits (discussed in our contributions).

In \texttt{SynNonIID} we have two users, each training an SVM ($f_1,f_2$, solid line) on 2-dimensional data points $x_1,x_2$ of one class exclusively (cf. \Cref{fig:intution_convergence_e,fig:intution_convergence_f}). The dotted line illustrates the margin: in the SVM training objective function $\mathcal{J}_{\text{HingeSVM}}$, we sum the hinge losses $\ell_\mathrm{hinge}$ over all data points where only those within this margin---the so-called support vectors---have a non-zero loss. The blindly averaged SVM (green), i.e. $f = 0.5 \cdot (f_1 + f_2)$, is the same as a central SVM (black), i.e., trained on the combined point cloud (cf. \Cref{fig:intution_convergence_g}). The reason why the averaged SVM is vertical, although the local SVMs are horizontal, is explained by the fact that the parameters of an SVM are normal vectors to the hyperplane and the local normal vectors point in the opposite direction (indicated by the arrow and the different background colors).

~

\denseparagraph{Problem Statement 1.} Prior work \citep{smith2017interaction} has shown an impossibility result for general non-interactive learning methods with strong utility-privacy tradeoffs. Other prior work on non-interactive DP convex learning \cite{daniely2019locally,dagan2020interaction} identified that the error introduced by non-interactivity also depends on the size of the margin, yet they focus on the LDP model and the learning of Boolean functions. \citet{jayaraman2018distributed} has shown that the utility from the local training is preserved which leaves a significant gap of $\sfrac{1}{w}$ to centralized learning. These results leave the question open: Can we understand the utility of \blindavg more precisely, i.e., when it works and when it fails?

~

\denseparagraph{Our Contribution 1.} We answer three questions:
\begin{enumerate}[leftmargin=*,label=(C1.\arabic*)]
    \item \emph{Is there a criterion that can guide when \blindavg will succeed and when it will fail?}
    In \Cref{fig:blindaveragingvsregularization} we empirically identify the L2-regularization parameter $\Lambda$ (a.k.a. strong convexity constant) of the objective $\mathcal{J}$ as the key utility driver:
    \blindavg works well with datasets that are robust against $\Lambda$, i.e., work with larger regularization (e.g. \texttt{SynNonIID} with $\Lambda=20$), and has issues with datasets that only perform with small regularization (e.g. \texttt{SynFail} with $\Lambda=0.05$). This limitation is similar to DP ERM learners where less noise is used for a larger $\Lambda$ as the sensitivity scales inversely with $\Lambda$ (cf. \Cref{thm:sensitivity_general}).

    \item \emph{Can we characterize the utility properties of \blindavg?}
    We use the following property: a hinge-loss SVM $f$ is uniquely defined by its support vectors $x_j$ (with $\alpha_j \neq 0$) following the representer theorem (cf. \Cref{thm:representer}) which states that ERMs admit the so-called dual form $f=\sum_{j=1}^n \alpha_j x_j$ with dual coefficients $\alpha_j$. We show:

    \begin{lemma}\label{lem:supportvec}
        A locally trained hinge-loss linear SVM has the support vectors $V^{(i)} \subseteq D^{(i)}$. Then, the average $\frac{1}{w}\sum_{i=1}^w f^{(i)}$ has the support vectors $V = \bigcup_{i=1}^{w} V^{(i)}$.
    \end{lemma}

    If we increase the margin by increasing $\Lambda$, all data points are support vectors (e.g. as in \texttt{SynNonIID}). Hence, with the same support vectors, both SVMs are the same and converge (cf. \Cref{cor:convergenceOfBlindlyAveraging}). In detail, we can upper bound the error of \blindavg with the quantitative and qualitative difference between the dual coefficients $\alpha_j$ of the local and central SVMs. For instance, \texttt{SynFail} exhibits a high error bound, as the dual coefficients significantly differ.

    \item \emph{What is the privacy-utility tradeoff of blind averaging compared to our baseline GradAvg+SecSum for a varying number of users?}
    We evaluate \blindavg of SVMs (called \ppmlname) on three datasets (cf. \Cref{fig:all_exps_highlevel}) and find that it approaches a utility comparable to our baseline on CIFAR-10 but struggles on CIFAR-100 and Federated EMNIST with significantly more classes.
    In an ablation study, the utility costs of \blindavg towards the centralized setting is even in an extreme non-IID scenario small.

\end{enumerate}

\subsection{A Novel Convex ERM Learner for \blindavg: \softmaxslp}
\label{sec:intro_softmaxreg}
Before introducing \softmaxslp, we recall prior results of a related learner: DP SVM.
The privacy-preservation of \blindavg is characterized via Differential Privacy (DP) \citep{Dwork2006} which has been shown to work well in the context of machine learning algorithms. DP quantifies the protection that any person has in a given learner against an attacker who can observe the resulting model and has arbitrary background knowledge.

\begin{definition}[Differential Privacy]
    Let $\mathrm{Obs}$ be a set of observations, and $\RV(\mathrm{Obs})$ be the set of random variables over $\mathrm{Obs}$, and $\mathcal{D}$ be the set of all datasets.
    A randomized algorithm $M\colon \mathcal{D} \to \RV(\mathrm{Obs})$ for all pairs of datasets $D,D' \in \mathcal{D}$ that differ in at most 1 element is a $(\eps, \delta)$-DP mechanism if for all tests $S\subseteq \mathrm{Obs}$: $\Pr[M(D)\in S] \le \exp(\eps) \Pr[M(D')\in S] + \delta$ with $\eps, \delta \in \mathbb{R}_+$.
\end{definition}

The so-called output sensitivity is a sufficient assumption for many DP mechanisms and ensures a bound on the influence of each individual's data on the resulting ML model.\footnote{For our proofs, we utilize a randomized variant of the sensitivity as proposed by \citet{wu2017bolt}.}

\begin{definition}[Sensitivity]
\label{def:sensitivity}
    Let $q\colon (D,r) \to \mathbb{R}$ be a randomized function on dataset $D$ and randomness $r$.
    The \emph{sensitivity} of $q$ is defined as $s = \max_{D\sim_1 D'}\max_r\lVert q(D,r) - q(D',r) \rVert$, where $D \sim_1 D'$ denotes that the datasets $D$ and $D'$ differ in at most one element.
\end{definition}

We follow the so-called output perturbation paradigm: each user locally noises their model $f^{(i)} = T_\xi(D^{(i)})$ obtained after locally training on dataset $D^{(i)}$ with algorithm $T_\xi$ and hyperparameters $\xi$: $f_{\mathrm{priv}} = T_\xi(D^{(i)}) + \mathcal{N}(0,\Sigma)$. If $T_\xi$ is $s$-sensitivity bounded, i.e. $s = \max_{D \sim_1 D'} \max_r \lVert T_\xi(D,r) - T_\xi(D',r) \rVert$, then $T_\xi$ is $(\eps,\delta)$-DP (cf. \Cref{cor:erm_is_dp}) by the Gaussian mechanism such that $\eps \in \mathcal{O}(s \cdot \sqrt{K_{\mathrm{comp}}})$ with $K_{\mathrm{comp}}$ as the number of compositions (e.g. the number of classes in an SVM).

For a class of strongly convex machine learning algorithms $T_\xi$ such as SVMs, \citet{chaudhuri2011differentially,wu2017bolt} have shown output sensitivity bounds. \citet{wu2017bolt} uses the concept of uniform stability \citep{hardt2016train} to derive bounds for each iteration of an SGD-based training. Their result requires three properties of the objective function $\mathcal{J}$ in $T_\xi$ to hold:

\begin{itemize}
    \item For all parameters $f$ and $f'$, the function $q$ is \textbf{$\Lambda$-strongly convex} if
    $\textstyle q(f) \le q(f') + \langle\nabla q(f'),f-f'\rangle + \frac{\Lambda}{2} \norm{f - f'}^2$. Simply put, strong convexity lower bounds the second derivative of $q$.

    \item For all parameters $f$ and $f'$, function $q$ is \textbf{$L$-Lipschitz continuous} if
    $\textstyle \frac{\norm{q(f) - q(f')}}{\norm{f - f'}} \le L$. Simply put, the Lipschitzness upper bounds the first derivative of $q$.

    \item For all parameters $f$ and $f'$, function $q$ is \textbf{$\beta$-smooth} if
    $\textstyle \frac{\norm{\nabla_f q(f) - \nabla_{f'}q(f')}}{\norm{f - f'}} \le \beta$. Simply put, smoothness upper bounds the second derivative of $q$.
\end{itemize}

\begin{theorem}[Lemma 8 in \citet{wu2017bolt}]
\label{thm:sensitivity_general}
    If a learner $T_\xi$ is trained on dataset $D$ with SGD on learning rate $\tau_m = \min(\sfrac{1}{\beta}, \sfrac{1}{\Lambda m})$ for iteration $m$ and has a $\Lambda$-strongly convex, $\beta$-smooth, and $L$-Lipschitz objective function $\mathcal{J}$, then the output model $f = T_\xi(D)$ has the sensitivity bound $s = \sfrac{2L}{\Lambda n}$.
\end{theorem}

One learner for which these SGD-bounds are known is the SVM learner with a Huber loss on relaxation parameter $h$ (cf.~\Cref{app:huber})---essentially a smoothed hinge-loss.

\begin{example}[$T=\text{\ppmlname}$]
    \label{ex:dpsvm}
    Assume a $c$-bounded input space $\mathcal{X}$, i.e. $\forall x\in\mathcal{X}\colon \norm{x} \le c$, and a $p$-dimensional $R$-bounded model parameter space $\mathcal{F}$ in each training iteration $m$, i.e. $\forall f_m\in\mathcal{F}\colon \norm{f_m} \le R$.
    The objective function of Huber loss SVM training (cf. \Cref{alg:dpsvm}) is $\Lambda$-strongly convex, ($L = \Lambda R + c$)-Lipschitz, and $(\beta=\sqrt{p\Lambda^2 + (\sfrac{c^2}{2h} + \Lambda)^2} \approx \sqrt{p}\Lambda)$-smooth. Thus, SGD-based SVM training (\ppmlname) has a sensitivity of $s = \frac{2(\Lambda R + c)}{\Lambda n}$.
\end{example}

Current learners that satisfy output sensitivity are limited to binary-class learners like SVMs or logistic regression (LR), although in a multi-class setting, these learners exhibit a disadvantage in the utility-privacy-tradeoff. To emulate a multi-class model, SVMs frequently use the one-vs-rest (OVR) scheme where $K$-many SVMs are learned such that each of the $K$ classes is trained against all others.
The OVR scheme has ($K_\mathrm{comp} = K$)-many compositions and does not offer a good selection process of the most likely class which merely is an argmax over all binary SVM predictions.

~

\denseparagraph{Problem Statement 2.}
Can we design another DP learner that satisfies output perturbation and solves the multi-class overhead of SVMs, i.e., uses no composition ($K_\mathrm{comp} = 1$) and has a good selection process of the most likely class?

~

\denseparagraph{Our Contribution 2.} We propose to use a single-layer softmax-activated perceptron (\softmaxslp) in the multi-class setting which, in contrast to SVMs, optimizes the class selection via the softmax cross-entropy loss. Due to output perturbation, \softmaxslp learning does not involve composition ($K_\mathrm{comp} = 1$).
We decisively answer the following three questions:  
\begin{enumerate}[leftmargin=*,label=(C2.\arabic*)]
    \item \emph{Does \softmaxslp satisfy output sensitivity?}
    We show output sensitivity and thus DP for SGD-trained \softmaxslp (\softmaxname) by proving that its training objective satisfies the requirements from \Cref{thm:sensitivity_general} \citep{wu2017bolt}. While it is known that the regularized \softmaxslp is $\Lambda$-strongly convex and the unregularized \softmaxslp is Lipschitz, we additionally show that the regularized \softmaxslp is ($\beta=\sqrt{(p+1)K \Lambda^2 + 0.5(\Lambda + c^2)^2} \approx \sqrt{p\cdot K} \Lambda$)-smooth (cf. \Cref{thm:smooth_softmax}) and ($L = \Lambda R + \sqrt2 c$)-Lipschitz (cf. \Cref{thm:lipschitz_softmax}). Thus:
    \begin{theorem}
    \label{thm:sens_softmax}
        $T=\text{\softmaxname}$ (cf. \Cref{alg:dpsoftmax}) has a sensitivity of $s = \frac{2(\Lambda R + \sqrt2 c)}{\Lambda n}$.
    \end{theorem}

    \item \emph{How does the privacy-utility tradeoff differ between the two learners: \softmaxname and OVR-trained \ppmlname?}
    Both learners differ in the 3 parameters: $\beta$, $L$, and $K_{\mathrm{comp}}$.
    \begin{itemize}[leftmargin=*]
        \item The $\beta$-smoothness differs by about a factor of $\sqrt{K}$ which leads to a smaller upper bound of the learning rate of \softmaxname and thus a possibly longer training time.
        
        \item The $L$-Lipschitzness is proportional to the sensitivity and differs only by a constant ($\sqrt{2}$). Hence, \softmaxname has a sensitivity independent of the number of classes $K$ for a constant $R$-bounded model space.
        
        \item The number of compositions $K_{\mathrm{comp}}$ differs by a factor of $K$ which leads to a $\sqrt{K}$-times higher privacy budget $\eps$ of \ppmlname than for \softmaxname as $\eps$ increases by roughly $\sqrt{K_{\mathrm{comp}}}$.\footnote{An OVR-trained SVM still has the utility disadvantage with the class selection above \softmaxslp. For a fairer comparison, we can consider the one-vs-one (OVO) scheme where $\Theta(K^2)$-many SVMs are learned such that each class is trained against each other. OVO has ($K_\mathrm{comp} = \Theta(K^2)$)-many compositions, which further increases $\eps$ by $\sqrt{K}$.}
        Often, the model size scales with the number of classes $K$ which would scale the maximal norm of the parameters $R$ by $\sqrt{K}$. In this case, \softmaxname still saves a privacy budget of $\frac{\eps_{\mathrm{svm}}}{\eps_{\mathrm{softmax}}} \propto \frac{\sqrt{K}\cdot(\Lambda \frac{R}{\sqrt{K}} + c)}{\Lambda R + \sqrt{2}c} = \frac{\Lambda R + \sqrt{K}c}{\Lambda R + \sqrt{2}c}$.
    \end{itemize}

    \item \emph{Does \softmaxslp outperform SVMs and achieve competitive performance on real-world datasets?} 
    While our experiments in \Cref{fig:all_exps_highlevel} show that the utility gap between \softmaxname and our highly interactive baseline persists on CIFAR-100 and Federated EMNIST, \softmaxname almost entirely closes the gap on CIFAR-10. \softmaxname also outperforms \ppmlname on all three multi-class datasets. 
    Our ablation shows that the utility cost of \blindavg compared to the centralized setting is small in an extreme non-IID scenario when the noise is low, e.g., for a large dataset. In particular, our experiments show that \ppmlname is more robust to adding noise than \softmaxname.
\end{enumerate}

\section{Preliminaries}
    The following details \Cref{ex:dpsvm} from the introduction:

\begin{algorithm}[!h]
   \caption{\Cref{ex:dpsvm}: $T_\xi=\text{\ppmlname{}}_\xi(D)$ with hyperparameters $\xi \coloneqq (K,c,M,h,\Lambda,R)$}
   \label{alg:dpsvm}
\begin{algorithmic}
    \STATE {\bfseries Input:} dataset $D \coloneqq \myset{(x_j,y_j)}_{j=1}^n$ where data point $x_j$ {{\parfillskip0pt\par}}
    \begin{ALC@g}
        \STATE is structured as $[1, x_{j,1}, \dots, x_{j,p}]$; \#classes $K$; input clipping bound: $c \in \mathbb{R}_+$; \#iterations $M$; Huber loss relaxation $h \in \mathbb{R}_+$; regularization parameter: $\Lambda \in \mathbb{R}_+$; model clipping bound: $R \in \mathbb{R}_{+}$
    \end{ALC@g}
    \STATE {\bfseries Result:} $\{\, f_M^{(k)} \,\}_{k\in \myset{1,\dots,K}} \in \mathbb{R}^{(p+1) \times K}$: models with hyperplanes $\in \mathbb{R}^p$ and intercepts $\in \mathbb{R}$
    \STATE $\operatorname{clipped}(x) \coloneqq {c\cdot {x}}/{\max(c, \norm{x})}$
    \STATE $\mathcal{J}(f, D, k) \coloneqq \textstyle  \frac{1}{n}\sum_{(x,y)\in D} \ell_{\mathrm{huber}}\big(h, y \langle f,\operatorname{clipped}(x)\rangle \cdot (1[y=k] - 1[y\neq k])\big) + \frac{\Lambda}{2} \langle f,f \rangle$
    \FOR{$k$ {\bfseries in} $1,\dots,K$}
    \FOR{$m$ {\bfseries in} $1,\dots,M$}
    \STATE $f_m^{(k)} \gets f_{m-1}^{(k)} - \tau_m \nabla\mathcal{J}(f_{m-1}^{(k)},(x_j,y_j),k)$, with learning rate $\tau_m = \min(\frac{1}{\beta}, \frac{1}{\Lambda m})$,
    \begin{ALC@g}
        \STATE $\beta = \sqrt{(\sfrac{c^2}{2h} + \Lambda)^2 + p\Lambda^2}$, and index $j=m \operatorname{mod} n$
    \end{ALC@g}
    \STATE $f_m^{(k)} \gets R \cdot f_m^{(k)}/\lVert f_m^{(k)}\rVert$ \hfill\COMMENT{projected SGD}
    \ENDFOR
    \ENDFOR
\end{algorithmic}
\end{algorithm}

    \begin{table*}[!h]
    \caption{Our Notation.}
  \label{tab:configuration}\label{def:conf}
  \centering
  \begin{tabular}{llllll}
    \toprule
    Symbol & Description & Symbol & Description & Symbol & Description \\
    \cmidrule(r){1-2} \cmidrule(l){3-4} \cmidrule(l){5-6}
    $U^{(i)} \in \mathcal{U}$ & $i$-th out of $w$-many users & $T_\xi(D)$ & $s$-sensitivity bounded learner & $K_{\mathrm{comp}}$ & number of compositions \\
    $t$ & ratio of honest users & & e.g. \ppmlname, \softmaxslp & $K$ & number of classes \\
    $D^{(i)} \subseteq \mho$ & local dataset of the $i$-th user &  $\xi$ & hyperparameters of $T$ & $M$ & number of training iterations \\
    $n^{(i)} $ & number of local data points & $\avg(T)$ & blind model average: $\frac{1}{w} \sum_{i=1}^{w} T_\xi(D^{(i)})$ & $c,R$ & input \& model clipping bound \\
    $(x,y)$ & label $y$, data point $x \in \mathbb{R}^p$ & $f$ or $\alpha$ & model parameter; dual: $f = \sum_{j=1}^n \alpha_j x_j$ & $L,\beta,\Lambda$ & $L$-Lipschitz, $\beta$-smooth, $\Lambda$-regularized \\
    $V \subseteq D$ & set of support vectors & $\mathcal{J}$ & objective function on inputs $(f,D)$ & $\sigma \in \mathbb{R}_+$ & noise scale \\
    \bottomrule
  \end{tabular}
\end{table*}

\subsection{Differential Privacy}
\label{app:dpdefs}
    \denseparagraph{Computational Differential Privacy}
Note that because of the secure summation, we technically require the computational version of differential privacy \citep{mironov2009computational}, where the differential privacy guarantees are defined against computationally bounded attackers; the resulting increase in $\delta$ is negligible and arguments about computationally bounded attackers are omitted to simplify readability.

\begin{definition}[Computational $\approx_{\eps,\delta}^c$ Differential Privacy]
\label{def:comp-dp-rel}
    Let $\mathcal{D}$ be the set of all datasets and $\eta$ a security parameter.
    Given a randomized algorithm $M\colon \mathcal{D} \to \RV(\mathrm{Obs})$ and a pair of datasets $D,D' \in \mathcal{D}$, we write $M(D) \approx_{\eps,\delta}^c M(D')$ if for any polynomial-time probabilistic attacker
	    $\Pr[A(M(D)) = 0] \le \exp(\eps) \Pr[A(M(D')) = 1] + \delta(\eta)$.
    For all pairs of datasets $D,D'$ that differ in at most $1$ element $M$ is a computational $(\eps, \delta(\eta))$-DP mechanism if we have
	    $M(D) \approx_{\eps,\delta}^c M(D')$.
\end{definition}

\subsection{SGD-trained ERMs are Differentially Private}

Differential privacy for an $s$-sensitivity bounded learner $T_\xi$ directly follows from differential privacy of the Gaussian mechanism.

\begin{definition}[Delta-Gaussian]
\label{def:deltagauss}
    Given noise scale $\Tilde{\sigma} \in \mathbb{R}_+$, complementary error function $\erfc$, $\sigma_{\mathrm{p}} \coloneqq \sfrac{1}{\Tilde{\sigma}}$, and $\mu_{\mathrm{p}} \coloneqq \sfrac{\sigma_{\mathrm{p}}^2}{2}$, we define $\delta(\eps, K_{\mathrm{comp}}) = 0.5 \cdot ( \erfc(\frac{ \eps - K_{\mathrm{comp}}\mu_{\mathrm{p}} }{ \sqrt{2K_{\mathrm{comp}}} \sigma_{\mathrm{p}} }) - e^{\eps} \erfc(\frac{ \eps + K_{\mathrm{comp}}\mu_{\mathrm{p}} }{ \sqrt{2K_{\mathrm{comp}}} \sigma_{\mathrm{p}} }) )$.
\end{definition}

\begin{lemma}[Gaussian mechanism is DP, Theorem 5 in \citet{sommer2019privacy} \& \Cref{lem:tight_adp_gauss}]
\label{lem:gauss_mech}
    Let $q_k$ be $s$-sensitivity-bounded functions on dataset $D$ and $\mathcal{N}$ a multivariate Gaussian. The Gaussian mechanism $D \mapsto \{ q_k(D) + \mathcal{N}(0,\Tilde{\sigma}^2 s^2 I) \}_{k \in \{1,\dots,K_{\mathrm{comp}}\}}$ is $(\eps,\delta)$-DP with $\delta(\eps, K_{\mathrm{comp}})$ as in \Cref{def:deltagauss}. 
\end{lemma}

\begin{corollary}[Gaussian mechanism on $T_\xi$ is DP]
\label{cor:erm_is_dp}
    For $s$-sensitivity bounded learner $T_\xi$, $D \mapsto T_\xi(D) + \mathcal{N}(0,\Tilde{\sigma}^2 s^2 I_{(p+1) \times K})$ is $(\eps,\delta)$-DP with $\delta(\eps, K_{\mathrm{comp}})$ as in \Cref{def:deltagauss}.\footnote{By \Cref{thm:m2} in \Cref{app:randomized_sens} we can apply the Gaussian mechanism for a deterministic sensitivity (cf. \Cref{lem:gauss_mech}) to learner $T_\xi$ of \Cref{thm:sensitivity_general} that has a randomized sensitivity as in \Cref{def:sensitivity}.}
\end{corollary}

In essence, $\eps \in \mathcal{O}(s \cdot \sqrt{K_{\mathrm{comp}}})$. \citet{balle2020privacy} have shown a similar tight composition result.

\subsection{Secure Summation}
\label{app:securesum}
    Hiding intermediary local training results as well as ensuring their integrity is provided by an instance of secure multi-party computation (MPC) called secure summation \citep{bonawitz2017practical,bell2020secure}. It is targeted to comply with distributed summations across a huge number of parties. In fact, \citet{bell2020secure} has a computational complexity for $w$ users on an $l$-sized input of $\mathcal{O}(\log^2 w + l\log w)$ for the user and $\mathcal{O}(w(\log^2 w +l\log w))$ for the server as well as a communication complexity of $\mathcal{O}(\log^2 w + l)$ for the user and $\mathcal{O}(w(\log w + l))$ for the server thus enabling an efficient run-through of roughly $10^9$ users without biasing towards computationally equipped users. Additionally, it offers resilience against user dropouts and colluding adversaries, both of which are substantial features for our distributed setting.

\begin{definition}[Secure Summation]
\label{def:secagg}
    Let $\mathcal{F}(s_1,\dots,s_n) \coloneqq \sum_{i=1}^w s_i$. We say that $\pi_{\mathrm{SecSum}}$ is secure summation if there is a probabilistic polynomial-time simulator $\mathrm{Sim}_{\mathcal{F}}$ such that if a fraction of users is corrupted ($C \subseteq \myset{U^{(1)},\dots, U^{(w)}}$, $|C| = \gamma w$), $\mathrm{Real}_{\pi_{\mathrm{SecSum}}}(s_1$, $\dots$, $s_w)$ is statistically indistinguishable from $\mathrm{Sim}_{\mathcal{F}}(C, \mathcal{F}(s_1,\dots,s_w))$, i.e., for an unbounded attacker $\mathcal{A}$ there is a negligible function $\nu$ such that
    \begin{align*}
        \operatorname{Advantage}&(\mathcal{A}) = \lvert\Pr[
        \langle\mathcal{A},\mathrm{Real}_{\pi_{SecAgg}}(s_1,\dots,s_w)\rangle = 1] - \\
        &\Pr[\langle\mathcal{A},\mathrm{Sim}_{\mathcal{F}}(C, \mathcal{F}(s_1,\dots,s_w))\rangle = 1]\rvert \le \nu(\eta).
    \end{align*}
    Here, $\mathrm{Sim}_{\mathcal{F}}$ is a potentially interactive simulator that only has access to the sum of all elements and the (sub)-set of corrupted users. The adversary is unable to distinguish interactions and outputs of the simulator from those of the real protocol.
    For a detailed definition of the network execution $\mathrm{Real}_{\pi}$ using the notion of interactive machines, we refer to \Cref{app:extended_secsum}.
\end{definition}

The following theorem is proven for global network attackers that are passive and statically compromised parties. Formally, the theorem holds for all attackers $(\mathcal{A'}, \mathcal{A''})$ of the following form. $\mathcal{A'}$ internally runs $A''$ and ensures that only static compromization is possible and that the attacker remains passive. 

\begin{theorem}[Secure Aggregation $\pi_{SecAgg}$ in the semi-honest setting exists \citep{bell2020secure}]
\label{theorem:secure_aggregation}
    Let $s_1, \dots, s_n$ be the $d$-dimensional inputs of the users $U^{(1)}, \dots, U^{(w)}$. Let $\mathcal{F}$ be the ideal secure summation function: $\mathcal{F}(s_1, \dots, s_n) \coloneqq \sum_{i=1}^w s_i$. If secure authentication encryption schemes and authenticated key agreement protocol exist, the fraction of dropouts (i.e., users that abort the protocol) is at most $\rho \in [0,1]$, at most a $\gamma \in [0,1]$ fraction of users is corrupted ($C \subseteq \set{U^{(1)},\dots, U^{(w)}}$, $|C| = \gamma w$), and the aggregator is honest-but-curious, then there is a secure summation protocol $\pi_{SecAgg}$ for a central aggregator and $w$ users that securely emulates $\mathcal{F}$ as in \Cref{def:secagg}.
\end{theorem}

\begin{table*}[!ht]
    \caption{
        \textbf{Comparison to related work} for $w$ users with $n$ data points each and $M$ training iterations: utility-costs, i.e., convergence to the centralized setting before noise, DP noise scale, and number of secure summation (SecSum) invocations.
        (\checkmark) denotes experimental evidence without formal proof and $^\ast$ denotes communication rounds without invoking SecSum.
    }
  \label{tab:realted_work_comp}
  \centering
  \begin{tabular}{llllr}
    \toprule
    SVM Algorithms & Utility costs & DP Noise & Invocations \\
    \midrule
    \quad \citet{jayaraman2018distributed}, gradient perturbation & \checkmark & $\mathcal{O}(\sfrac{\sqrt M}{nw})$ & $\mathcal{O}(\log(nw))$ \\
    \quad \citet{jayaraman2018distributed}, output perturbation & (\checkmark) & ${\mathcal{O}(\sfrac{1}{nw})}$ & ${1}$ \\
    \quad \textbf{\blindavg: SVM (ours)} & \checkmark$^\text{(\Cref{cor:convergenceOfBlindlyAveraging})}$ & ${\mathcal{O}(\sfrac{1}{nw})}$ & ${1}$ \\
    \midrule
    \softmaxslp Algorithms &  &  & \\
   \midrule
    \quad GradAvg & \checkmark & $\mathcal{O}(\sfrac{\sqrt M}{(n\sqrt{w})})$ & $M^\ast$ \\
    \quad GradAvg + SecSum \cite{AgSuYuKuMa_18:cpsgd,kairouz2021distributed} & \checkmark & $\mathcal{O}(\sfrac{\sqrt M}{nw})$ & $M$ \\
    \quad \textbf{\blindavg: \softmaxslp (ours)} & (\checkmark) & ${\mathcal{O}(\sfrac{1}{nw})}^\text{(\Cref{thm:sens_softmax})}$ & ${1}$ \\
    \midrule
    Baseline: Centralized training & \checkmark & ${\mathcal{O}(\sfrac{1}{nw})}$ & ${0}$ \\
    \bottomrule
  \end{tabular}
\end{table*}

\subsection{Dual SVM Representation}
\label{app:representer_theorem}
    With the representer theorem, we can completely describe a converged SVM $T_\xi(D)$ on an $n$-sized dataset $D$ using an $\alpha_j$-weighted linear combination of the data points: $T_\xi(D) = \sum_{j=1}^n \alpha_j x_j$.
This idea of a new representation of $T_\xi(D)$ is at the core of the kernel method and the dual problem formulation of SVMs.
In general, the representer theorem assumes an ERM objective with a specific regularizer like the L2-regularizer:

\begin{theorem}[Representer theorem, cf. Lemma 3 and Theorem 8 in \citet{argyriou2009there}]
\label{thm:representer}
    Given a dataset $D \coloneqq \set{(x_j,y_j)}_{j=1}^n \subseteq \mathcal{H} \times \mathcal{Y}$ on a Hilbert space $\mathcal{H}$ with $\dim(\mathcal{H}) \ge 2$ and label space $\mathcal{Y}$, and a trained model $f$ of learner $T_\xi$ such that there exists a solution that belongs to $\spn(\set{x_j}_{j=1}^n)$, where $f = T_\xi(D) = \argmin_{f \in \mathcal{H}} E(\set{\inner{f}{x_j}, y_j}_{j=1}^n) + \Lambda\Omega(f)$ for some arbitrary error function $E\colon (\mathbb{R} \times \mathcal{Y})^n \to \mathbb{R}$ and $\Lambda$-factorized differentiable regularizer $\Omega\colon \mathcal{H} \to \mathbb{R}$. Then $T_\xi$ admits a solution of the form
    $f = \sum_{j=1}^{n} \alpha_j x_j$ for some $\alpha_j \in \mathbb{R}$ if and only if $\forall{f \in \mathcal{H}}\colon \Omega(f) = h(\inner{f}{f})$ with $h\colon \mathbb{R}_+ \to \mathbb{R}$ as a non-descreasing function.
\end{theorem}

In the case of \ppmlname and \softmaxname, we have $\Omega = \inner{f}{f}$ which fulfills the requirements of the representer theorem since $h(z) = z$ is a linear function and the learner $T$ follows the definitions after convergence: $E(\set{\inner{f}{x_j}, y_j}_{j=1}^n) = \frac{1}{n} \sum_{(x,y) \in D^{(i)}} \ell_{\mathrm{huber}}(y \inner{f}{x})$ is the error function of \ppmlname and $E(\set{\inner{f}{x_j}, y_j}_{j=1}^n) = \frac{1}{n} \sum_{(x,y) \in D^{(i)}} \ell_{\mathrm{softmax}}(y \inner{f}{x})$ the one of \softmaxname.

\section{Related Work}
\label{section:related-work}
    Here and in \Cref{tab:realted_work_comp}, we discuss the most related work.
For interactive ERM via gradient perturbation, \citet{jayaraman2018distributed} have shown strong utility-privacy tradeoffs. It requires $\mathcal{O}(\log(nw))$ secure summation (SecSum) invocations for $w$ users with $n$ data points each.
For non-interactive learning, however, \citet{jayaraman2018distributed} have shown output perturbation results using \citet{chaudhuri2011differentially} for which they only showed that the convergence bound from the local training is preserved, thus leaving a gap of $\sfrac{1}{w}$ to centralized learning.
We show that there exists a regularization constant $\Lambda$ such that SVM learning converges, closing the $\sfrac{1}{w}$ gap, and we experimentally show that \softmaxslp leads to strong results. Moreover, their work does not exclude leakage from the learner (only from the optimum), whereas we use bounds on the learner SGD~\citep{wu2017bolt}.

In differentially-private gradient averaging (GradAvg, also called federated learning) \citep{McMahanMRHA17,abadi2016deep,chen2022fundamental,choquette2024amplified}, each user submits noisy local model updates protected via DP guarantees (DP-SGD). A central server then aggregates all incoming updates. 
However, GradAvg's noise scales with $\mathcal{O}(\sqrt{w})$ and is prohibitively interactive as its communication rounds increase with the number of training iterations. 
Recent work~\citep{AgSuYuKuMa_18:cpsgd,kairouz2021distributed} aggregates the local model updates securely via secure summation (`GradAvg + SecSum') while adding the same amount of noise as in centralized training.
Although it can utilize one of the more efficient MPC protocols \cite{bonawitz2017practical,bell2020secure} and can handle user fluctuations and dropouts, it is non-interactive as it requires one MPC invocation per training iteration.
GradAvg variants \cite{McMahanMRHA17} that perform more than one local model update to reduce the number of shared model updates are also interactive. Its non-interactive variant of just one shared update step is very similar to \blindavg: If the $i$-th user shares $\sfrac{1}{w}\cdot(f^{(i)}_\mathrm{trained} - f_\mathrm{init})$ then we effectively blindly average the locally trained models $f^{(i)}_\mathrm{trained}$: $\sfrac{1}{w}\sum_{i=1}^w (f^{(i)}_\mathrm{trained} - f_\mathrm{init}) = \avg(f^{(i)}_\mathrm{trained}) - f_\mathrm{init}$.

~

\denseparagraph{Other Privacy-preserving Distributed Learning Protocols.}
The noise overhead of GradAvg can be completely avoided by protocols that rely on cryptographic methods to hide intermediary training updates from a central aggregator. Several secure distributed learning methods protect the contributions during training but do not come with privacy guarantees for the model such as DP: an attacker, e.g. a curious training party, can potentially extract information about the training data from the model. As we focus on differentially private distributed learning methods, we will neglect those methods.

cpSGD \citep{AgSuYuKuMa_18:cpsgd} is a protocol that utilizes secure multi-party computation (MPC) methods to honestly generate noise and compute DP-SGD. While cpSGD provides the full flexibility of SGD, it does not scale to millions of users as it relies on expensive MPC methods. \citet{TrBaAnStLuZhZh_19:hybrid} relies on a combination of MPC and DP methods which also does not scale to millions of users.

Another line of research aims for the stronger privacy goal of protecting a user's entire input, called local DP, during distributed learning \citep{BaKaMaThTh_2020:checkins,GiDaDiKaSu_21:shuffle_fl}. Due to the strong privacy goal, gradient averaging with local DP tends to achieve weaker accuracy. With \Cref{cor:group_privacy}, evaluated in \Cref{fig:exp_heatmap}, we show how \blindavg achieves a comparable guarantee via group privacy and MPC-based aggregation overhead: given enough users, any user can protect their entire dataset at once while we still reach good accuracy.

For DP SVM training, other methods besides output perturbation \citep{chaudhuri2011differentially,wu2017bolt} such as objective perturbation \citep{chaudhuri2011differentially,kifer2012private,iyengar2019towards,bassily2019private} and gradient perturbation \citep{bassily2014private, wang2017differentially, feldman2018privacy, bassily2019private, feldman2020private,yu2020gradient} exist.
Output perturbation noises a trained model once calibrated to the model's sensitivity which marks it an ideal candidate for non-interactive learning. 
In contrast, objective perturbation noises the objective function, while gradient perturbation noises the intermediary gradient updates. Yet, DP requires objective perturbation and the privacy amplification by iteration variant of gradient perturbation \citep{feldman2018privacy,feldman2020private} to leak no intermediary model updates which marks in the convex SVM setting their amplification above DP-SGD. To protect all of these updates in a distributed setting, the training has to be performed in expensive protocols like MPC or homomorphic encryption. Other gradient perturbation methods that leak intermediary gradient \citep{bassily2014private,abadi2016deep,yu2020gradient} need one MPC invocation per iteration. All these variants have a large communication overhead and are thus highly interactive.

~

\denseparagraph{Trustworthy Distributed Noise Generation.}
One core requirement of MPC-based distributed learning is honestly generated and unleakable noise, as otherwise, our privacy guarantees would not hold anymore.
There is a rich body of work on distributed noise generation \citep{moran2009optimally,dwork2006our, kairouz2015secure, kairouz2021distributed, goryczka2015comprehensive}. So far, however, no distributed noise generation protocol scales to millions of users.
Thus, we use a simple, yet effective technique: we add enough noise if at least a fraction of them (say $\noncolluding = 50\,\%$) are not colluding to violate privacy by sharing the noise they generate with each other.

\section{Our Contribution 1: Formal Utility Guarantees of \blindavg}
\label{sec:noninteractive}
    We derive a convergence of blindly averaging a hinge-loss linear SVM $T=\text{HingeSVM}$ for some regularization constant $\Lambda$ to centralized training and identify the difference between an averaged and a central SVM in the respective support vectors and, more precisely, the dual coefficients $\alpha_i$.

By the representer theorem (cf. \Cref{thm:representer}~\citep{argyriou2009there}), ERMs like a locally trained SVMs admit the dual form $f^{(i)} = \sum_{j=1}^n \alpha_j x_j$ with $x_j \in D^{(i)}$ as a local data point. Since the local datasets $D^{(i)}$ are disjoint, our \Cref{lem:averagerepresenter} (proof: \Cref{app:proof_averagerepresenter}) shows that the average of ERMs has the union of the local dual coefficients as dual coefficients. Thus, if only a few $\alpha_j$'s differ between the averaged local models and the central model, in the worst case, the error of \blindavg is significantly smaller as if a lot $\alpha_j$'s differ. If all $\alpha_j$ are the same, \blindavg converges.

\begin{corollary}[Averaged Representer theorem]
    \label{lem:averagerepresenter}
    If a local learner $T_\xi$ on dataset $D^{(i)}$ admits a solution of the form $f^{(i)} = T_\xi(D^{(i)}) = \sum_{j=1}^{n} \alpha_j^{(i)} x_j^{(i)}$ (cf. \Cref{thm:representer}) then the average $f \coloneqq \avg(f^{(i)})$ admits a solution of the form $f = T_\xi(\mho) = \frac{1}{w}\sum_{i=1}^{w}\sum_{j=1}^{n} \alpha_{j}^{(i)} x_{j}^{(i)}$ with $\mho \coloneqq \bigcup_{i=1}^w D^{(i)}$.

\end{corollary}

For hinge-loss SVMs ($f^{(i)} \coloneqq \argmin_{f} \frac{1}{n} \sum_{(x,y) \in D^{(i)}} \max(0,1-y \inner{f}{x})\allowbreak + \Lambda\inner{f}{f}$), \Cref{lem:averagerepresenter} implies \Cref{lem:supportvec} (proof: \Cref{app:proof_supportvec}) that we announced in the introduction: if each converged local HingeSVM with the same local data sizes, i.e. $\forall{i,i'}\colon n^{(i)} = n^{(i')}$, has support vectors $V^{(i)}$ then its average has support vectors $V = \bigcup_{i=1}^{w} V^{(i)}$.

\begin{replemma}{lem:supportvec}
    A locally trained hinge-loss linear SVM has the support vectors $V^{(i)} \subseteq D^{(i)}$. Then, the average $\frac{1}{w}\sum_{i=1}^w f^{(i)}$ has the support vectors $V = \bigcup_{i=1}^{w} V^{(i)}$.

\end{replemma}

If the support vectors and thus all $\alpha_j$ of an averaged and a central SVM are the same, we converge (cf. \Cref{cor:convergenceOfBlindlyAveraging}, proof: \Cref{app:proof_avg_converges}).
Such a scenario occurs e.g. if the regularization $\Lambda$ is high and thus the margin is large enough such that all data points are within the margin, i.e., support vectors: $V = \mho$. For formal convergence, we use the SGD-training variant $T=\text{HingeSVM-SGDWA}$ that uses weighted averages to reach a convergence rate of $\mathcal{O}(\sfrac{1}{M})$ with the number of iterations. Other SGD training variants may achieve a slower convergence rate.

\begin{theorem}[Averaging locally trained SVMs converges to a central SVM]
	\label{cor:convergenceOfBlindlyAveraging}
    Given the same local data sizes $\forall{i,i'}\colon n^{(i)} = n^{(i')}$, there exists a regularization parameter $\Lambda$ such that the average of HingeSVMs trained with projected subgradient descent using weighted averaging $T=\text{HingeSVM-SGDWA}$ converges with the number of local iterations $M$ to the best central model with $\mho \coloneqq \bigcup_{i=1}^w D^{(i)}$, i.e.
$\Eop[\mathcal{J}(\avg(\text{HingeSVM-SGDWA}), \mho) - \inf_{f} \mathcal{J}(f,\mho)] \in \mathcal{O}(\sfrac{1}{M})$.

\end{theorem}

While this non-private \blindavg converges, a differentially private variant naturally retains an additive error due to the added noise on the model parameters $f$ after finishing training. In particular, the error introduced by additive Gaussian noise $\mathcal{N}(0,\Tilde\sigma)$ on $f$ amounts to
$\Eop[\lvert(f + \mathcal{N}(0,\Tilde\sigma)) - f\rvert] =  \Eop[\lvert \mathcal{N}(0,\Tilde\sigma) \rvert] = \Tilde\sigma\sqrt{\sfrac{2}{\pi}}$.

\section{Our Contribution 2: Differentially Private SoftmaxReg for Output Perturbation}
\label{sec:softmax}
    \begin{algorithm*}[!h]
   \caption{Our $T_\xi=\text{\softmaxname{}}_\xi(D)$ with hyperparameters $\xi \coloneqq (K,c,M,\Lambda,R)$}
   \label{alg:dpsoftmax}
    \begin{algorithmic}
        \STATE {\bfseries Input:} dataset $D \coloneqq \{(x_j,y_j)\}_{j=1}^n$ where data point $x_j$ is structured as $[1, x_{j,1}, \dots, x_{j,p}]$; \#classes $K$; {{\parfillskip0pt\par}}
        \begin{ALC@g}
             \STATE input clipping bound: $c \in \mathbb{R}_+$; \#iterations $M$; regularization parameter: $\Lambda \in \mathbb{R}_+$; model clipping bound: $R \in \mathbb{R}_{+}$
        \end{ALC@g}
        \STATE {\bfseries Result:} $f_M \in \mathbb{R}^{(p+1) \times K}$: a model with hyperplane $\in \mathbb{R}^{p \times K}$ and intercept $\in \mathbb{R}^K$

        \STATE $\operatorname{clipped}(x) \coloneqq {c\cdot {x}}/{\max(c, \norm{x})}$
        \STATE $\mathcal{J}_{\mathrm{softmax}}(f,D) \coloneqq \frac\Lambda2 \sum_{k=1}^K \langle f_k, f_k\rangle + \frac1n\sum_{(\operatorname{clipped}(x),y)\in D} -\sum_{k=1}^K y_k\log\frac{\exp\langle f_k, x\rangle}{\sum_{j=1}^K \exp\langle f_j, x\rangle}$
        \FOR{$m$ {\bfseries in} $1,\dots,M$}
        \STATE $f_m \gets f_{m-1} - \tau_m \nabla\mathcal{J}_{\mathrm{softmax}}(f_{m-1},(x_{j},y_{j}),k)$, with learning rate $\tau_m \coloneqq \min(\frac{1}{\beta}, \frac{1}{\Lambda m})$,
        \begin{ALC@g}
            \STATE $\beta = \sqrt{(d+1)K \Lambda^2 + 0.5(\Lambda + c^2)^2}$, and index $j=m \operatorname{mod} n$.
        \end{ALC@g}
        \STATE $f_m \gets R \cdot f_m/\lVert f_m\rVert$ \hfill\COMMENT{projected SGD}
        \ENDFOR
    \end{algorithmic}
\end{algorithm*}

We show the output sensitivity bound for \softmaxname (cf. \Cref{alg:dpsoftmax}) that we announced in the introduction:

\begin{reptheorem}{thm:sens_softmax}
    $T=\text{\softmaxname}$ (cf. \Cref{alg:dpsoftmax}) has a sensitivity of $s = \sfrac{2(\Lambda R + \sqrt2 c)}{\Lambda n}$.
\end{reptheorem}

Differential privacy of \softmaxname (cf. \Cref{cor:softmax_is_dp}) directly follows from a bounded sensitivity (cf. \Cref{thm:sens_softmax}) and DP of the Gaussian mechanism (cf. \Cref{cor:erm_is_dp}).

\begin{corollary}
\label{cor:softmax_is_dp}
    For an $s$-sensitivity-bounded learner $T=\text{\softmaxname}$ (cf. \Cref{thm:sens_softmax}), $D \mapsto \text{\softmaxname{}}_\xi(D)\allowbreak + \mathcal{N}(0,\Tilde{\sigma}^2 s^2 I_{(p+1)\times K})$ is $(\eps,\delta)$-DP with $\delta(\eps,K_{\mathrm{comp}} = 1)$ as in \Cref{def:deltagauss}.
\end{corollary}

To prove \Cref{thm:sens_softmax}, we use the output sensitivity requirements in \Cref{thm:sensitivity_general} \citep{wu2017bolt} which states that, besides the bounded learning rate, it suffices to show:

\begin{theorem}[simplified]
\label{thm:strongconvexity_softmax}
    The objective function $\mathcal{J}_{\mathrm{softmax}}$ is $\Lambda$-strongly convex.
\end{theorem}

\begin{theorem}[simplified]
    \label{thm:lipschitz_softmax}
    The objective function $\mathcal{J}_{\mathrm{softmax}}$ is ($L = \Lambda R + \sqrt2 c$)-Lipschitz.
\end{theorem}

\begin{theorem}[simplified]
    \label{thm:smooth_softmax}
    The objective function $\mathcal{J}_{\mathrm{softmax}}$ is ($\beta=\sqrt{(p+1)K \Lambda^2 + 0.5(\Lambda + c^2)^2} \approx \sqrt{p\cdot K} \Lambda$)-smooth.
\end{theorem}

We now describe the details of these proofs and refer to \Cref{app:strongconvexity_softmax} for the full theorem statements and proofs.

~

\denseparagraph{Lipschitzness.}
The sensitivity of \softmaxname and thus the privacy budget $\eps$ is directly proportional to its Lipschitzness $L = \Lambda R + \sqrt{2}c$. The constant $L$ is proven in \citet[Appendix D]{das2023beyond} for a non-regularized objective, and we prove it in \Cref{app:lipschitz_softmax} for a regularized one which shows to be independent of the number of classes for a fixed $R$ ($\forall f\colon \inner{f}{f} \le R$). The proof bounds the Jacobian of the objective $\mathcal{J}_{\mathrm{softmax}}$, i.e. $\sup_{z\in D, f}\lVert\nabla_{f}\mathcal{J}_{\mathrm{softmax}}(f,z)\rVert \le L$. The $\Lambda R$-part of the bound $L$ originates from the L2-regularization term in the objective, i.e., $\sfrac{\Lambda}{2}\inner{f}{f}$, which is influenced by the size of model $f$ whereas the $\sqrt{2} c$-part originates from the softmax cross-entropy loss for which we use the characteristic of the softmax that the probabilities of each class add up to $1$. In particular, we use the fact that for the $K$ softmax probabilities $s_1,\dots, s_K$ and class label $y$: $\max_{s_1,\dots, s_K} \{\, (\sum_{k=1}^K (s_k - 1_{[y=k]})^2)^{\sfrac{1}{2}} \mid \sum_{k=1}^K s_k = 1 \land \forall k\colon s_k \ge 0 \,\} = \sqrt{2}$. $L$ contains this $\sqrt{2}$.

~

\denseparagraph{Smoothness.}
By \citet[Theorem 1]{zhou2018fenchel}, smoothness and convexity of the primal problem imply a strongly convex dual problem. The smoothness is also used as an upper bound on the learning rate in \softmaxname.
We show in \Cref{app:smooth_softmax} the smoothness $\beta=\sqrt{(p+1)K \Lambda^2 + 0.5(\Lambda + c^2)^2}$ by bounding the Hessian of the objective: $\sup_{z\in D, f} \lVert \mathbf{H}_{f}(\mathcal{J}_{\mathrm{softmax}}(f,z)) \rVert \le \beta$.  The first part of the smoothness bound $\beta$, $(p+1)K\Lambda^2$, stems from the L2-regularization term of the objective function $\sfrac{\Lambda}{2}\langle f, f \rangle$ which is influenced by the size of model $f \in \mathbb{R}^{(p+1)\times K}$. In particular, the second derivative of the regularization term is constant in each direction of the derivative, thus marking the dependence on the number of model parameters $(p+1)K$. The second part of $\beta$, $0.5(\Lambda + c^2)^2$, stems from the softmax cross-entropy loss for which we use the characteristic of the softmax that the probabilities of each class add up to $1$. In particular, we also use the fact that $\max_{s_1,\dots,s_K} \{\, \sum_{k=1}^K s_k(1 - s_k)(C + s_k) \mid \sum_{k=1}^K s_p = 1 \land \forall k\colon s_k \ge 0 \,\} \le 0.25(C+1)^2$ for $C \propto \sfrac{\Lambda}{c^2}$ which we prove in \Cref{lem:max_smoothness} using the KKT conditions. The bound $0.25(C+1)^2$ scales proportional to $2c^4$ (cf. \Cref{thm:smooth_softmax}) which directly corresponds to the $0.5(\Lambda + c^2)^2$ term in $\beta$.

~

\denseparagraph{Strong Convexity.}
The strong convexity parameter $\Lambda$ stems from the regularization term $\sfrac{\Lambda}{2}\inner{f}{f}$ and we use and show in \Cref{app:strongconvexity_softmax} that the objective function without the regularization is convex. In particular, the Hessian of the softmax-activated cross-entropy loss function $\mathcal{L}_{\mathrm{CE}}(y, z) \coloneqq -\sum_{k=1}^K y_k\log\frac{\exp z_k}{\sum_{j=1}^K \exp z_j}$ is convex if it is positive semi-definite: $\nabla^2\mathcal{L}_{\mathrm{CE}} \succeq 0$.

\section{Setup of \blindavg}
\label{sec:setup}
    In our evaluation, we use an implementation of \blindavg (cf. \Cref{alg:dist_blindavg}) which follows the scheme of \citet{jayaraman2018distributed} with two extensions: 1) a threat model with dishonest users (cf. \Cref{sec:threat_model}) and 2) a parameter upscaling to accommodate differing local data sizes $n^{(i)}$ (cf. \Cref{sec:parameter_upscaling}). As a secure summation protocol, we use a non-interactive extension of the SecAgg protocol \citep{bell2020secure} (cf. \Cref{sec:non_interactive_secsum}) and abstractly notate its user and server part in \Cref{alg:dist_blindavg} with $\pi_{\mathrm{SecSum}}^{(\mathrm{User})}$ and $\pi_{\mathrm{SecSum}}^{(\mathrm{Server})}$. SecAgg requires $4$ communication rounds, but the user only shares their local model in the third round.
Following \citet{bogetoft2009secure}, we introduce $J$ computation servers that aggregate the model parameters on behalf of the users.
To ensure that our variant of \blindavg preserves privacy, we show:

\renewcommand{\algorithmicdo}{}
\begin{algorithm}[H]
    \caption{
        \blindavg{}. $\pi_{\mathrm{SecSum}}$ as in \Cref{def:secagg}.
    }
    \label{alg:dist_blindavg}
    \begin{algorithmic}
        \WHILE{{User \blindavg{}}($D^{(i)}$, $w$, t, $\sigma$, T, $\xi$)}
            \STATE {\bfseries Input:} local dataset $D^{(i)}$ with $n^{(i)} = \abs{D^{(i)}}$; \#users $w$; {{\parfillskip0pt\par}}
            \begin{ALC@g}
                 \STATE  ratio $\noncolluding$ of honest users; noise scale $\sigma$; learner $T$; hyperparameters $\xi$ incl. \#classes $K$
            \end{ALC@g}
            \STATE {\bfseries Result:} $\textstyle f_{\mathrm{priv}} \in \mathbb{R}^{(p+1) \times K}$: DP-models
            \STATE $f \gets T_\xi(D^{(i)})$ \hfill\COMMENT{$T$ is $s$-sensitivity-bounded}
            \STATE $f_{\mathrm{priv}} \gets f + \mathcal{N}(0,\Tilde{\sigma}^2  s^2 I_{(p+1)\times K})$
            \begin{ALC@g}
                \STATE with $\Tilde{\sigma} \coloneqq \sigma\cdot \sfrac{1}{\sqrt{\noncolluding\cdot w}}$
            \end{ALC@g}
            \STATE $\pi_{\mathrm{SecSum}}^{(\mathrm{User})}(\sfrac{n^{(i)}}{w} \cdot f_{\mathrm{priv}})$\;
        \ENDWHILE
        \item[]
        \WHILE{{Server \blindavg{}}($\mathcal{U}$)}
            \STATE {\bfseries Input:} users $\mathcal{U}$;
            \STATE {\bfseries Result:} empty string
            \STATE $\pi_{\mathrm{SecSum}}^{(\mathrm{Server})}(\mathcal{U})$
        \ENDWHILE
    \end{algorithmic}
\end{algorithm}
\renewcommand{\algorithmicdo}{\textbf{do}}

\begin{theorem}[simplified]
\label{thm:distributed_blindavg}
    \blindavg{} (cf. \Cref{alg:dist_blindavg}) satisfies computational $(\eps, \delta+\nu)$-DP for all neighboring central datasets $\mho,\mho'$ with $\delta(\eps,K_{\mathrm{comp}})$ as in \Cref{def:deltagauss} and a function $\nu$ negligible in the security parameter used in $\pi_{\mathrm{SecSum}}$.
\end{theorem}

Simplified, the proof follows by applying the sensitivity after \blindavg (cf.~\Cref{cor:ampl_avg}) to the Gaussian mechanism (cf.~\Cref{lem:gauss_mech}) where the noise is applied per user (cf.~\Cref{thm:sqrtgauss}). Since we average the noisy local models, we also average the noise. This has the effect that it suffices if each user only adds noise of scale $\tilde\sigma = \sfrac{1}{\sqrt{\noncolluding\cdot w}}$ instead of $\tilde\sigma = \sfrac{1}{\noncolluding\cdot w}$.
The detailed privacy analysis is in \Cref{sec:security_distblindavg}.

\subsection{Threat Model and its Implications on our Experiments}
\label{sec:threat_model}
We assume that a fraction of at least $t$ users are honest (say $t=50\,\%$), i.e., they follow the protocol including honestly generated noise and do not collude with the adversary. In contrast, untrustworthy users can collude with a passive, collaborating adversary by exchanging information about the randomness used in their local computation.
The adversary is assumed to have full knowledge about each user's dataset, except for one data point of one user. 
To compensate for untrustworthy users, we adjust the noise added by each user by $t$; e.g., if $t = 50\,\%$, then we double the noise to satisfy our guarantees.

\subsection{Parameter Upscaling}
\label{sec:parameter_upscaling}
Before aggregating the local models, we upscale the model by the number of local data points $n^{(i)}$. This upscaling ensures that all local sensitivities are the same and independent of $n^{(i)}$ and thus allows differentially private blind averaging with differing $n^{(i)}$. Utility-wise, the prediction of SVMs and \softmaxslp are scale-invariant, i.e., the prediction is the same if we scale the model by any constant. If all $n^{(i)}$ are the same, this upscaling corresponds to an averaged model scaled by a constant $n$ which has no utility implications due to this scale-invariant nature.

\subsection{Non-interactive SecSum Protocol}
\label{sec:non_interactive_secsum}
\blindavg is agnostic to a specific secure summation protocol $\pi_{\mathrm{SecSum}}$ (cf.~\Cref{alg:dist_blindavg}): e.g. we can use a non-interactive extension of the SecAgg protocol \citep{bell2020secure}. SecAgg requires $4$ communication rounds, but the user only shares their local model in the third round.
Following \citet{bogetoft2009secure}, we introduce $J$ computation servers that aggregate the model parameters on behalf of the users. Specifically, each user $i$ sends their model $f^{(i)}$ in fixed-point arithmetic in shares $r^{(j,i)}$ to server $j$, where for $j < J$, each $r^{(j,i)}$ is drawn randomly from $\myset{1, \ldots, B}$ for a sufficiently large $B$ and where $r^{(J,i)} = (f^{(i)}- \sum_{j<J} r^{(j,i)})\mod B$. The computation servers then run SecAgg among each other, yielding the sum of all inputs. All $r^{(j,i)}$ cancel out and the sum over all models $f^{(i)}$ remains. The secret sharing technique is information-theoretically secure if at least one computation server is honest. Security assumptions of SecAgg apply to the computation servers instead of the users. 
Although secure, this protocol is not robust against active attacks.

\begin{figure*}[!ht]
    \centering
    \subfigure[Dataset: CIFAR-10]{{\resizebox{0.325\textwidth}{!}{\input{plots/baseline_dist_always_all_data_1000only.pgf}\unskip}}}
    \subfigure[Dataset: CIFAR-100]{{\resizebox{0.325\textwidth}{!}{\input{plots/baseline_dist_always_all_data_cifar100_100only.pgf}\unskip}}}
    \subfigure[Dataset: Federated EMNIST]{{\resizebox{0.325\textwidth}{!}{\input{plots/baseline_dist_always_all_data_emnist.pgf}\unskip}}}

    \caption{
        \textbf{Main result} (detailed plot: \Cref{fig:all_exps}). Classification accuracy vs. $\eps$ of \blindavg ($\delta = 10^{-5}$, $t=50\,\%$ honest users) on \softmaxslp (cf. \Cref{alg:dpsoftmax}) and SVM (cf. \Cref{alg:dpsvm}), DP-SGD-based gradient averaging (GradAvg), and GradAvg+SecSum (as a baseline). Both GradAvg variants are highly interactive. \ppmlname underperformed for CIFAR-100 and is below the plotting range. Since \blindavg assumes dishonest users, it uses twice as much noise.
    }
    \label{fig:all_exps_highlevel}
\end{figure*}

\begin{figure*}[!h]
    \centering
    \subfigure[HingeSVM: SynNonIID]{{\resizebox{0.245\textwidth}{!}{\input{plots/gen_convergance_intuition_5.pgf}\unskip}}}
    \subfigure[HingeSVM: SynFail]{{\resizebox{0.245\textwidth}{!}{\input{plots/gen_convergance_intuition2_5.pgf}\unskip}}}
    \subfigure[CIFAR-10]{{\resizebox{0.245\textwidth}{!}{\input{plots/gen_convergance_intuition_CIFAR10.pgf}\unskip}}}
    \subfigure[CIFAR-100]{{\resizebox{0.245\textwidth}{!}{\input{plots/gen_convergance_intuition_CIFAR100.pgf}\unskip}}}
    
    \caption{
        \textbf{Accuracy for varying regularization parameters $\Lambda$} of \blindavg on four datasets.
    }
    \label{fig:blindaveragingvsregularization}
\end{figure*}

\subsection{Security of \blindavg}
\label{sec:security_distblindavg}
    First, we derive a tight output sensitivity bound. A naïve approach would be to release each individual predictor, determine the noise scale proportionally to $\Tilde{\sigma} \coloneqq \sigma$ (cf. \Cref{cor:erm_is_dp}), showing $(\eps,\delta)$-DP for every user. We can save a factor of $\sqrt{w}$ by leveraging that $w$ is known to the adversary and we have at least $\noncolluding = 50\,\%$. Consequently, local noise of scale $\Tilde{\sigma} \coloneqq \sigma\cdot \sfrac{1}{\sqrt{\noncolluding\cdot w}}$ is sufficient for $(\eps,\delta)$-DP.

\begin{lemma}[Privacy amplification via averaging]
\label{cor:ampl_avg}
    With the notation in \Cref{def:conf}, \blindavg{} of \Cref{alg:dist_blindavg} without noise, $\avg(n^{(i)} \cdot T_\xi(D^{(i)}))$, has a sensitivity of $s' \cdot \sfrac{1}{w}$ for each model if $s = \sfrac{s'}{n}$.

\end{lemma}

The proof is in \Cref{sec:proof_ampl_avg}. The sensitivity of the aggregate is bounded to $s' \cdot \sfrac{1}{w}$ by rescaling the local models by $n^{(i)}$ which leads to local sensitivities independent of $n^{(i)}$ and allows blind averaging with varying local data sizes $n^{(i)}$. The sensitivities of $T = \text{\ppmlname}$ (cf.~\Cref{ex:dpsvm}) and $T = \text{\softmaxname}$ (cf. \Cref{thm:sens_softmax}) fulfill the condition in the Lemma as they are proportional to $n^{-1}$, thus: $s' = s \cdot n$.
Next, we show that locally adding noise per user $\Tilde{\sigma}$ proportional to $\sigma \cdot \sfrac{n^{(i)}}{\sqrt{w}}$ and taking the mean over the users is equivalent to centrally adding noise $\Tilde{\sigma}$ proportional to $\sigma \cdot \sfrac{n^{(i)}}{w}$.
Adding dishonest noise can be treated as post-processing and does not impact privacy.

\begin{lemma}
\label{thm:sqrtgauss}
    With the notation in \Cref{def:conf} and noise scale $\Tilde{\sigma}$:
$\frac{1}{w}\sum_{i=1}^{w} \mathcal{N}(0,(\Tilde{\sigma}\cdot\sfrac{1}{\sqrt{w}})^2) = \mathcal{N}(0,(\Tilde{\sigma}\cdot\sfrac{1}{w})^2)$.

\end{lemma}

The proof is in \Cref{sec:proof_sqrtgauss}. We now prove differential privacy for \blindavg of \Cref{alg:dist_blindavg} with noise scale $\Tilde{\sigma} \coloneqq \sigma\cdot \sfrac{1}{\sqrt{\noncolluding\cdot w}}$ and thus $\eps \in \mathcal{O}(\sfrac{s'}{\noncolluding\cdot w} \cdot\sqrt{K_{\mathrm{comp}}})$.

\begin{reptheorem}{thm:distributed_blindavg}[simplified]
    \blindavg{} (cf. \Cref{alg:dist_blindavg}) satisfies computational $(\eps, \delta+\nu)$-DP for all neighboring central datasets $\mho,\mho'$ with $\delta(\eps,K_{\mathrm{comp}})$ as in \Cref{def:deltagauss} and a function $\nu$ negligible in the security parameter used in $\pi_{\mathrm{SecSum}}$.
\end{reptheorem}

The full statement and proof are in \Cref{sec:blindavg_is_dp}. Simplified, the proof follows by applying the sensitivity (cf.~\Cref{cor:ampl_avg}) to the Gaussian mechanism (cf.~\Cref{lem:gauss_mech}) where the noise is applied per user (cf.~\Cref{thm:sqrtgauss}).

Next, we show how to protect the entire dataset of a single user (e.g., for distributed training via smartphones). The sensitivity-based bound on the Gaussian mechanism implies strong $\Upsilon$-group privacy results (see \Cref{sec:group_privacy}), leading to security guarantees as in local DP with the overhead of secure summation.

\begin{corollary}[Group-private variant]\label{cor:group_privacy}
    \blindavg{} of \Cref{alg:dist_blindavg} satisfies computational $(\Upsilon\eps, \delta + \nu)$, $\Upsilon$-group DP for all central datasets $\mho,\mho'$ differing in $\Upsilon$ many data points ($\Upsilon$-neighboring) with $\delta(\eps, K_{\mathrm{comp}})$ as in \Cref{def:deltagauss} and a function $\nu$ negligible in the security parameter used in $\pi_{\mathrm{SecSum}}$.

\end{corollary}

\blindavg uses local aggregators but can also protect all data points from a user with our user-level privacy generalization in \Cref{cor:userlevel_sens} based on \Cref{cor:group_privacy}.
For user-level privacy ($\forall i\colon \Upsilon = n^{(i)}$), it suffices that the norm of each model is bounded by $R$: then an averaged model has a sensitivity of $\sfrac{2R}{w}$.
The proof is in \Cref{sec:userlevel_sens}.
We can also conclude the same sensitivity if we exchange a whole user dataset $D^{(i)}$.

\begin{corollary}[User-level sensitivity]\label{cor:userlevel_sens}
    With the notation in \Cref{def:conf}, we say that learner $T_\xi$ is $R$-norm bounded if for any local dataset $D^{(i)}$: $\norm{T_\xi(D^{(i)})} \le R$. Any $R$-norm bounded $T_\xi$ has a deterministic sensitivity $s = 2R$.
Then, $D^{(i)} \mapsto T_\xi(D^{(i)})\allowbreak + \mathcal{N}(0,\Tilde{\sigma}^2 s^2 I_{p \times K})$ satisfies computational $(\Upsilon\eps,\delta)$, $\Upsilon$-group DP for all $\Upsilon$-neighboring central datasets $\mho,\mho'$ if $\forall i,j\colon n^{(i)} = n^{(j)}$ with $\delta(\eps, K_{\mathrm{comp}})$ as in \Cref{def:deltagauss}, $\Upsilon = n^{(i)} = \abs{D^{(i)}}$, and $\nu$ negligible in the security parameter used in $\pi_{\mathrm{SecSum}}$.

\end{corollary}

\section{Experimental Results}
\label{sec:results}
    \denseparagraph{Experimental Setup.}
Unless stated differently, we leveraged $5$-repeated $6$-fold stratified cross-validation for all differential privacy CIFAR experiments and $10$-repeated cross-validation on the pre-defined split for differential privacy EMNIST ones. We conducted a hyperparameter search across $\Lambda$ and $R$ for each evaluation setting and $\eps$ and reported the mean accuracy of the best hyperparameter configuration. We test between 2--16 hyperparameter configurations per setting. Privacy Accounting has been done with the privacy bucket \citep{meiser2018tight, sommer2019privacy} toolbox\footnote{accessible at \url{https://github.com/sommerda/privacybuckets}, MIT license} or, for Gaussians without subsampling, with \citet[Theorem 5]{sommer2019privacy} where both can be extended to multivariate Gaussians (cf. \Cref{app:multivariate_gaussian_as_univariate}).
We set DP $\delta = 10^{-5}$ if not stated otherwise, which is for CIFAR below $\sfrac{1}{nw}$ where $nw$ is the size of the combined local data. To emulate a distributed dataset for CIFAR-10/100, each algorithm spreads the dataset randomly among the users.
More details on our experimental setup are postponed to \Cref{app:experimental_setup}. We provide code for \ppmlname and \softmaxname at \url{https://github.com/kirschte/blindavg}.

~

\denseparagraph{Pretraining Enables Linear Classification Head.}
As proposed by \citet{tramer2021differentially}, we used a SimCLR pretrained model\footnote{accessible at \url{https://github.com/google-research/simclr}, Apache-2.0 license} \citep{chen2020big} on ImageNet ILSVRC-2012 \citep{ILSVRC15} to get an embedding of the local data (cf. \Cref{fig:tsne_cifar10} in \Cref{app:pretraining_viz} for an embedding view). Their evaluation shows that a linear classification head on the embedding already performs better than a supervised-only model of the same size: $83.1\,\%$ vs $80.5\,\%$ accuracy (ImageNet, $1000$ classes).
The pretrained model is a ResNet152 with selective kernels \citep{li2019selective} and a width multiplier of $3$ (overall: $795$M parameters) in the fine-tuned SimCLR variant.
For EMNIST, we inverted the image.

~

\denseparagraph{Sensitive Datasets.}
CIFAR-10, CIFAR-100 \citep{Krizhevsky09learningmultiple}, and federated EMNIST\footnote{ref: \url{https://tensorflow.org/federated/api_docs/python/tff/simulation/datasets/emnist}}\citep{cohen2017emnist,caldas2018leaf} act as our sensitive datasets; all after SimCLR pretraining. CIFAR is frequently used as a benchmark dataset in DP and EMNIST in distributed learning literature. Both CIFAR datasets consist of $60{,}000$ thumbnail-sized, colored images of $10$ or $100$ classes. Federated EMNIST consists of $\approx 750{,}000$ thumbnail-sized, grayscale images of $62$ classes and is annotated with $3{,}400$ user-partitions based on the author of the images: users have between $19$ and $465$ data points, on average $220 \pm 85$.

\subsection{Evaluation}
We compare \blindavg of \ppmlname and \softmaxname to DP-SGD-based 1-layer gradient averaging (GradAvg) and  GradAvg+SecSum to answer the remaining questions:

\emph{(C1.1) Is there a criterion that can guide when \blindavg will succeed and when it will fail?}
\Cref{fig:blindaveragingvsregularization} presents the L2-regularization criterion as a utility driver of \blindavg.
For HingeSVM and \softmaxslp, \blindavg converges to the global non-private model with increasing $\Lambda$ and all models maintain close task-performance for a mid-range regularization. \blindavg fails for tasks not robust under regularization (i.e., those that only perform well with a small $\Lambda$) like the deliberately designed \texttt{SynFail} (cf. \Cref{fig:intution_convergence}) but succeeds at robust tasks like \texttt{SynNonIID} (cf. \Cref{fig:intution_convergence}), CIFAR-10, or CIFAR-100.
In contrast to HingeSVM, \ppmlname does not converge without error as it uses a relaxation of the hinge loss: the Huber loss where non--support vectors can influence the model independent of the regularization as there is no hard cutoff at the margin like in a hinge-loss SVM.

\emph{(C1.3) \& (C2.3) What is the privacy-utility tradeoff of blind averaging compared to our baseline of GradAvg+SecSum for a varying number of users?}
We expect that (a) blindly averaged \softmaxslp outperforms blindly averaged SVMs (cf. \Cref{sec:softmax}) as the number of classes increases and (b) the gap between blindly averaged SVM and our baseline decreases as the number of users increases (cf. \Cref{section:related-work}):
(a) Our experiments in a high-user scenario (cf. \Cref{fig:all_exps_highlevel}) and in a centralized setting (cf. \Cref{fig:baseline}) support the hypothesis that the gap of \softmaxname towards \ppmlname becomes wider the more classes the dataset has.
(b) Our experiments show that the utility gap to our baseline for both \softmaxname and \ppmlname decreases the more users $w$ partake for the same data size (cf. \Cref{fig:all_exps}) and the more both data and user base increase (cf. \Cref{fig:exp_fixed_eps} in the appendix).
Note that the performance gap in blind averaging between $1$ and $100$ users is largely due to our assumption of $\noncolluding = 50\,\%$ dishonest users which scales the noise by $\sqrt{2}$;
$t=1$ deactivates this disadvantage.

\emph{Follow-up question: (C1.3) \& (C2.3) How robust is \blindavg{}'s utility if the local data is non-IID?}
We observe for strongly biased non-IID data (cf. \Cref{tab:noniid}) that on CIFAR-10, the utility decline of \ppmlname is small, whereas \softmaxname needs more users for a similar utility preservation since it is more sensitive to noise.
For real-world federated EMNIST with unbalanced local data sizes $n^{(i)}$, we observe a notable performance gap between the averaged (3,400 users) and global variant (1 user) in both learners (cf. \Cref{fig:all_exps} (d)). 
We weigh each local SVM by $n^{(i)}$ to achieve the constant sensitivity per user. This has a utility disadvantage against non-weighted averaging, visible in the experiments.

\begin{table}[!t]
  \caption{
    \textbf{Strongly biased non-IID experiments for blind averaging (\blindavg)} ($\eps=1.2$): each user has exclusive access to only one class. We report the non-IID accuracy and compare it to our regular experiments in percentage points (pp).
    As in \Cref{fig:exp_heatmap}, we extrapolate the accuracy on datasets 67 times larger using less noise.
  }
  \label{tab:noniid}
  \centering
  \begin{tabular}{lrll}
    \toprule
    &  & \multicolumn{2}{c}{Accuracy on dataset} \\
    \cmidrule{3-4}
    \blindavg variant & \multirow[t]{2}{*}{\shortstack[l]{dataset\\ multiplier}} & CIFAR-10 & CIFAR-100 \\
    \midrule
    \ppmlname & 1x & $85$\,\% ($-2$\,pp)  & -- \\
    \ppmlname & 67x & $87$\,\% ($-3$\,pp)  & -- \\
    \softmaxname & 1x & $42$\,\% ($-49$\,pp) & $1.6$\,\% ($-43$\,pp) \\
    \softmaxname & 67x & $88$\,\% ($-4$\,pp) & $58$\,\% ($-4$\,pp) \\
    \bottomrule
  \end{tabular}
\end{table}

\begin{figure}[!t]
    \centering
    \subfigure[roughly $w=200{,}000$ users and $\delta = 10^{-10}$]{{\resizebox{0.43\textwidth}{!}{\input{plots/baseline_grouppriv_heatmap.pgf}\unskip}}}
    \subfigure[roughly $w=20{,}000{,}000$ users and $\delta = 10^{-12}$]{{\resizebox{0.43\textwidth}{!}{\input{plots/baseline_grouppriv_heatmap_1B.pgf}\unskip}}}

    \caption{
        \textbf{Accuracy for local aggregators (group sizes $\Upsilon < 50$, cf. \Cref{cor:group_privacy}) and user-level privacy ($\Upsilon = 50$, cf. \Cref{cor:userlevel_sens}) in \blindavg} for the \softmaxname learner on CIFAR-10 data.
        We take the best accuracy for $\Upsilon$-group DP (protects $\Upsilon$ out of $n=50$ local data points) and user-level privacy (protects the entire user) which already for $\Upsilon = 2$ pivots to the latter. 
        To emulate more users and larger datasets, we interpolated the accuracy of $1{,}000$ users on $50$ data points each to a rescaled $\eps$-value ($\eps' \coloneqq \sfrac{1000\cdot\eps\cdot\Upsilon}{w}$, approximates the actual $\eps'$). Thus, we report pessimistic accuracies as the accuracy does not increase with the users; actually averaging over all users should perform better.
    }
    \label{fig:exp_heatmap}
\end{figure}

\begin{figure}[!t]
    \centering
    \centerline{\resizebox{0.86\columnwidth}{!}{\input{plots/baseline_dist_eps_0.5885.pgf}\unskip}}
    \caption{CIFAR-10 accuracy vs. \#users with $50$ data points per user for $(\eps, \delta) = (0.6, 10^{-5})$. GradAvg values (without SecSum) are interpolated.}
    \label{fig:exp_fixed_eps}
\end{figure}

\emph{Follow-up question: (C1.3) \& (C2.3) How does \blindavg perform for user-level privacy if scaled with significantly more users.}
\Cref{fig:exp_heatmap} presents this connection. Most notably, for ($\Upsilon \ge 2$)-group DP, we observe a user-level sensitivity of $\sfrac{2R}{w}$ (cf.~\Cref{cor:userlevel_sens}) to be mostly tighter than the data point dependent one by \citet{wu2017bolt}.

~

\denseparagraph{Computation Costs.} Extrapolating \citet[Table 2]{bell2020secure}, we need for \blindavg with model size $\ell \approx 100{,}000$ (CIFAR-10) and $1{,}000$ users $\le 0.2s$ user time and $40s$ server time.

\section{Limitations \& Discussion}
\label{sec:lim_signalnoise}
    For datasets that work best with small regularization $\Lambda$, like the deliberately designed \textsc{SynFail}, \blindavg fails to capture a strong task-performance. Yet, with privacy, even the central model works best with mid-range $\Lambda \ge 1$ as the noise scales with $\mathcal{O}(\sfrac1\Lambda)$. This points to a more general limitation of small $\Lambda$'s for DP ERMs.
For unfavorable datasets like our non-IID variant of CIFAR-10, blind averaging leads to a reduced signal-to-noise ratio, i.e., model parameters are smaller than in the centralized setting. This may explain why blind averaged \softmaxslp{} works better for less noise (cf. \Cref{tab:noniid}).
For unbalanced datasets, increasing $\Lambda$ too much to help convergence can lead to poor accuracy as the margin grows so much that not the same number of support vectors is chosen per label.

For future work, a deeper understanding of \blindavg's limitations and convergence for other learners like \softmaxslp would be interesting.
There could also be train-and-pass variants for \blindavg (and also gradient averaging), where we blindly average a model not across all users but a subset and then pass this model to the next user subset. From the user's view, this variant is also non-interactive.

\subsection{Detailed Limitations}

\denseparagraph{Distributional Shift Between the Public and Sensitive Dataset.}
For pretraining, we leverage contrastive learning. While very effective generally, it is susceptible to performance loss if the shape of the sensitive data used to train an SVM or \softmaxslp is significantly different from the shape of the initial public data.

~

\denseparagraph{Input Clipping.}
We require bounded input data for a bounded sensitivity. In many pretraining methods like SimCLR, no natural bound exists, thus we norm-clip the input data by a constant $c$.
To provide a data-independent $c$ on CIFAR-10 and EMNIST, $c$ is based on CIFAR-100 (here: $34.854$); its similar data distribution encompasses the output distribution of the pretraining reasonably well. For CIFAR-100, $c$ is based on CIFAR-10 (here: $34.157$).

~

\denseparagraph{Hyperparameter Search.}
In \ppmlname, we have two important hyperparameters that influence the noise scale: the regularization weight $\Lambda$ and the predictor radius $R$. In the noise scale subterm, $\sfrac{c}{\Lambda}+R$, the maximal predictor radius is naturally significantly smaller than $\sfrac{c}{\Lambda}$ due to the regularization penalty. Thus, a bad-tuned $R$ often does not have as large of a utility impact as a bad-tuned $\Lambda$. Estimating parameters for a fixed $\eps$ from public data is called hyperparameter freeness in prior work \citep{iyengar2019towards}. For the other $\eps$ values, we can estimate $\Lambda$ by fitting a (linear) curve on related public data (proposed by \citet{chaudhuri2011differentially}) or synthetic data (proposed by AMP-NT \citep{iyengar2019towards}) as smaller $\eps$ prefer a higher $\Lambda$ and vice versa.

~

\denseparagraph{Blind Averaging---Signal-to-noise Ratio.}
For unfavorable yet balanced local datasets, we identify as a main limitation of blind averaging a reduced signal-to-noise ratio: the model is not as large as in the centralized setting with respect to the sensitivity analysis. This effect would explain why our experiments show that blind averaged \softmaxslp works better with very little noise  (cf. \Cref{tab:noniid}).

For SVMs, our formal characterization of the effect of blind averaging enables us to describe its limitations more precisely. In summary, we see two effects that reduce the signal-to-noise ratio. One effect comes from the requirement of the SVM training that the model with the smallest norm shall be found that satisfies the soft-margin constraints of the training data points. The local SVM training has fewer data points and, thus, fewer constraints. Hence, unfavorable local data sets will lead to a smaller model. Another effect comes from the averaging itself. Unfavorable local data sets can lead to local models that point in very different directions. When averaging these models, their norm naturally decreases as for any two vectors $a,b \in \mathbb{R}^p$ we have $0.5 \lVert a + b \rVert_2 \le 0.5 (\lVert a \rVert_2 + \lVert b \rVert_2)$, and this discrepancy is larger the smaller the inner product is.

~

\denseparagraph{Blind Averaging---Unbalanced Data.}
Convergence holds if all data points are support vectors (SV) which implies a large margin. Yet a regularly trained SVM chooses roughly equally many SVs per class: by the dual problem, we have the constraint $y^T\alpha=0$ for labels $y_j \in \myset{-1,1}$ and dual coefficients $\alpha$. If we have an SV inside the margin, then $\alpha_j=\Lambda^{-1}$.
Hence, enlarging the margin such that all data points are SVs can lead to poor utility performance. Moreover, unbalanced local data can deteriorate the performance of blind averaging as observed in the EMNIST experiments (cf. \Cref{fig:all_exps} (d)) as we favor privacy, i.e., a constant sensitivity per user, above utility, i.e., optimal local scaling.

~

\denseparagraph{Active Attacks.}
Active attackers may deviate from the protocol or send maliciously construed local models. If the used secure summation protocol is resilient against active adversaries and can still guarantee that only the sum of the inputs is leaked, privacy is preserved. This follows from analyzing our algorithm for just the honest users and then leveraging the post-processing property of differential privacy. Secure summation protocols such as \citet{bell2020secure} leak partial sums under active attacks and will diminish the privacy offered by our work against such adversaries as well.

\section*{Acknowledgments}
This project has been partially funded by the BMBF project MLens (16KIS1260K).\\
We thank Sayan Mukherjee for pointing us to the convergence results on averaged LASSO models, and we thank Kfir Yehuda Levy for pointing us to convex optimization literature.

\bibliographystyle{IEEEtranN}
\bibliography{ref}

\newpage
\appendices
\crefalias{section}{appendix}
\crefalias{subsection}{subappendix}
\crefalias{subsubsection}{subsubappendix}
\crefalias{subsubsubsection}{subsubsubappendix}
\section{Overview of Our Appendix}

\startcontents
\printcontents{}{1}{}

\section{Extended Evaluation Figure}
We refer to \Cref{fig:all_exps}.

\section{Detailed Background}
\label{app:extended_prelims}
\label{app:pretraining_viz}

\subsection{Pretraining to boost DP Performance}
\label{app:pretraining}
    \begin{figure}[!h]
    \centering
    \centerline{\includegraphics[width=\columnwidth]{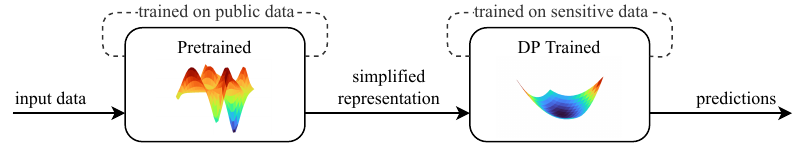}}
    \caption{Pretraining: Schematic overview. Dashed lines denote data flow during training and solid lines during inference.}
    \label{fig:schematic}
\end{figure}

\begin{figure}[!ht]
    \centering
    \centerline{\includegraphics[width=0.9\columnwidth]{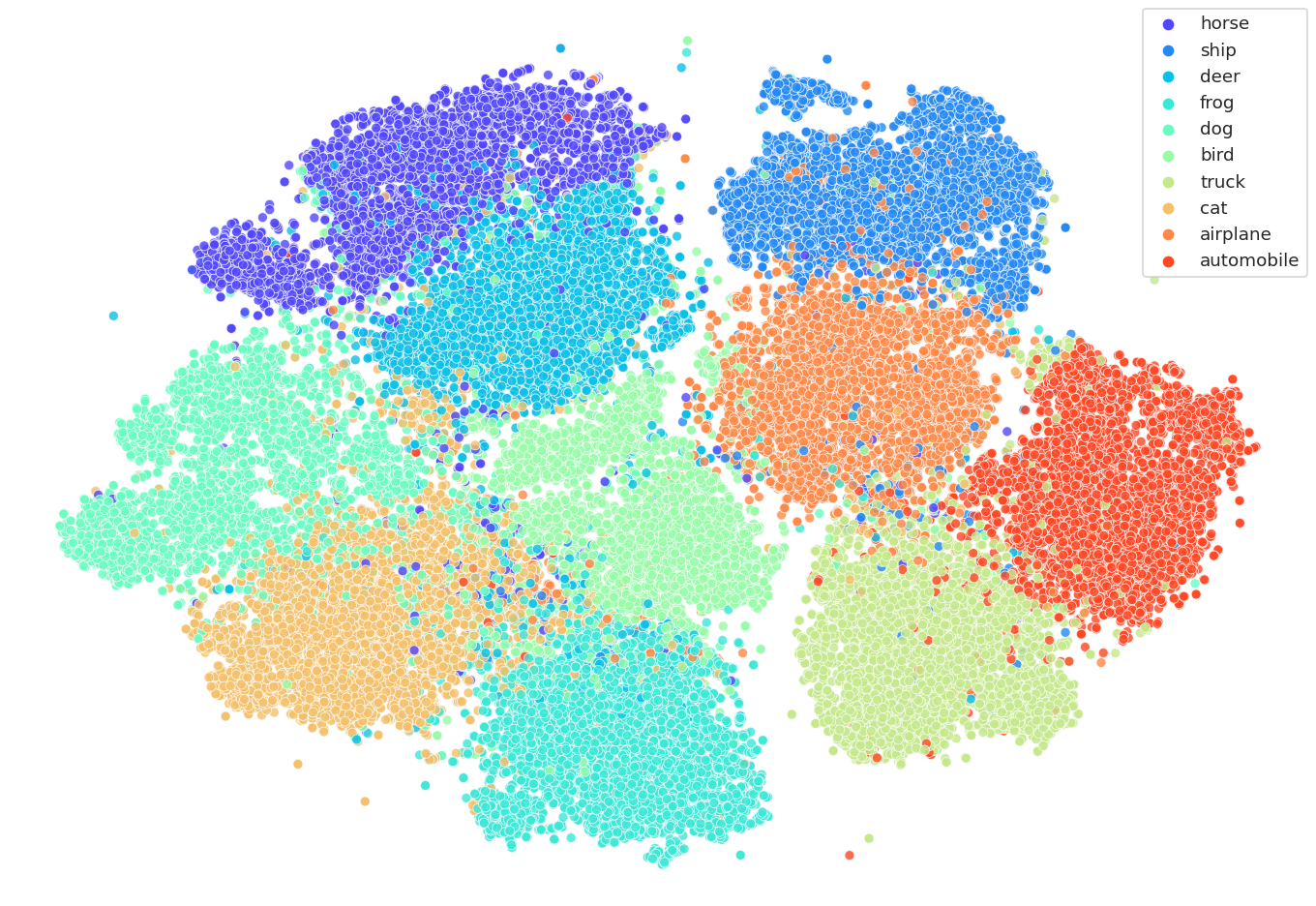}}
    \caption{$2$-d projection of the CIFAR-10 dataset via t-SNE \citep{van2008visualizing} with colored labels. t-SNE is defined on the local neighborhood, thus global structures may be arbitrary.}
    \label{fig:tsne_cifar10}
\end{figure}

\begin{figure*}[!ht]
    \centering
    \subfigure[Dataset: CIFAR-10]{{\resizebox{0.49\textwidth}{!}{\input{plots/baseline_dist_always_all_data.pgf}\unskip}}}
    \hfill
    \subfigure[Dataset: CIFAR-10]{{\resizebox{0.49\textwidth}{!}{\input{plots/baseline_dist.pgf}\unskip}}}
    \subfigure[Dataset: CIFAR-100]{{\resizebox{0.49\textwidth}{!}{\input{plots/baseline_dist_always_all_data_cifar100.pgf}\unskip}}}
    \hfill
    \subfigure[Dataset: Federated EMNIST]{{\resizebox{0.49\textwidth}{!}{\input{plots/baseline_dist_emnist.pgf}\unskip}}}

    \caption{
        \textbf{Classification accuracy} compared to privacy budget $\eps$ (in $\log$-scale) of \blindavg (cf.~\Cref{alg:dist_blindavg}), DP-SGD-based gradient averaging (GradAvg), and GradAvg + SecSum (always the 1 user line of GradAvg) ($\delta = 10^{-5}$). (b) Different numbers of users with 50 data points per user. (a,c,d) We use all available data points of the dataset for each line, spreading them among a differing number of users.
        For \blindavg and more than $1$ user, we assume honest noise of only $t=50\,\%$ of the users, thus scaling the local noise by a factor of $\sqrt{2}$.
        \ppmlname underperformed for CIFAR-100 and is below the plotting range.
    }
    \label{fig:all_exps}
\end{figure*}

Recent work \citep{tramer2021differentially, de2022unlocking} has shown that strong feature extractors (such as SimCLR \citep{chen2020simple, chen2020big}), trained in an unsupervised manner, can be combined with simple learners to achieve strong utility-privacy tradeoffs for high-dimensional data sources like images.
As a variation to transfer learning, it delineates a two-step process (cf. \Cref{fig:schematic}), where a simplified representation of the high-dimensional data (cf. \Cref{fig:tsne_cifar10} for a 2d visualization on CIFAR-10 data) is learned first before a tight privacy algorithm like \ppmlname or \softmaxname conducts the prediction process on these simplified representations. For that, two data sources are compulsory: a public data source used for a framework that learns a pertinent simplified representation and our sensitive data source that conducts the prediction process in a differentially private manner. Thereby, the sensitive dataset is protected while strong expressiveness is assured through the feature reduction network. Note that a homogeneous data distribution of the public and the sensitive data is not necessarily required.

Recent work has shown that for several applications, such representation reduction frameworks can be found, such as SimCLR for pictures, FaceNet for face images, UNet for segmentation, or GPT for language data. Without loss of generality, we focus in this work on the unsupervised SimCLR feature reduction network \citep{chen2020simple, chen2020big}. SimCLR uses contrastive loss and image transformations to align the embeddings of similar images while keeping those of dissimilar images separate \citep{chen2020simple}. It is based upon a self-supervised training scheme called contrastive loss where no labeled data is required. Labelless data is especially useful as it exhibits possibilities to include large-scale datasets that would otherwise be unattainable due to the labeling efforts needed.

\subsection{Huber Loss}
\label{app:huber}

\begin{definition}
    The Huber loss according to \citet[Equation 7]{chaudhuri2011differentially} is with a relaxation parameter $h$ defined as
    \begin{align*}
        \textstyle \ell_{\mathrm{huber}}(h, z) \coloneqq& \begin{cases} 0 &\mathrm{if } z > 1+h \\ \frac{1}{4h}(1+h-z)^2 &\mathrm{if } \abs{1-z} \le h\\ 1-z &\mathrm{if } z < 1-h \end{cases}.
    \end{align*}
\end{definition}

\subsection{Security of Secure Summation}
\label{app:extended_secsum}
    Before formulating the security of the secure summation protocol, we define a network execution against a global network attacker that is active and adaptive. For self-containedness, we briefly present the notion of interactive machines and a sequential activation network execution. More general frameworks for such a setting include, e.g., the universal composability framework~\citep{Ca00:uc}.

We rely on the notion of interactive machines.
For two interactive machines $X, Y$, we write $\langle X, Y \rangle$ for the interaction between $X$ and $Y$. We write $\langle X, Y \rangle = b$ to state that the machine $X$ terminates and outputs $b$.

~

\denseparagraph{The network execution $\mathrm{Real}_{\pi}$.}
Next, we define a network execution against a global network attacker that is active and adaptive. Given a protocol $\pi$ with user and server code, we define an interactive machine $\mathrm{Real}_{\pi}$ that lets each user party run the user code, lets the servers run the server code, and emulates a (sequential-activation-based) network execution, and interacts with another machine, called the attacker $\mathcal{A}$. The interaction is written as $\langle \mathcal{A}, \mathrm{Real}_{\pi}\rangle$.
Whenever within this network execution a party $B$ sends a message $m$ over the network to a party $C$, the interactive machine $\mathrm{Real}_{\pi}$, sends this message $m$ to the attacker, activates the attacker, and waits for a response $m'$ from the attacker. $\mathrm{Real}_{\pi}$ then lets this response $m'$ be delivered to party $C$, and activates party $C$. Moreover, the attacker $\mathcal{A}$ can send a dedicated message $(\texttt{compromise}, P)$ for compromising a party $P$ within the protocol execution. Whenever the attacker sends the message $(\texttt{compromise}, P)$ to the network execution $\mathrm{Real}_{\pi}$, the network execution marks this party $P$ as compromised and sends the internal state of this party to the attacker $\mathcal{A}$. For each compromised party $P$, the attacker decides how $P$ acts. Formally, the network execution redirects each message $m$ that is sent to $P$ to the attacker $\mathcal{A}$ and awaits a response message $(m', P')$ from the attacker $\mathcal{A}$. Upon receiving the response $(m',P')$, the network execution $\mathrm{Real}_{\pi}$ sends on behalf of $P$ the message $m'$ to the party $P'$.

For convenience, we write that a party $P$ \emph{runs the user code of a protocol $\pi$ on input $m$} when the network execution runs for party $P$ the user code of $\pi$ on input $m$.

\section{Extended Experimental Setup}
\label{app:experimental_setup}
    Concerning computation resources, our Python implementation of the EMNIST experiments with $3{,}400$ users took under a minute per user on a machine with 2x \emph{Intel Xeon Platinum 8168} @24\,Cores.

For \ppmlname-based experiments, we utilized projected SGD (PSGD) as used by \citet{wu2017bolt} and chose a batch size of $20$ and the Huber loss with relaxation parameter $h=0.1$.
For CIFAR-10, we chose a hypothesis space radius $R \in \myset{0.04, 0.05, 0.06, 0.07, 0.08}$, a regularization parameter $\Lambda \in \myset{10, 100, 200}$, and trained for $500$ epochs;
for the variant where we protect the whole local dataset, we chose $\Lambda \in \myset{0.5, 1, 2, 5}$ and $R \in \myset{0.06, 0.07}$ instead.
For CIFAR-100, we chose $R \in \myset{0.04, 0.06, 0.08}$, $\Lambda \in \myset{3, 10, 30, 100}$, and trained for $150$ epochs.
For EMNIST, we chose in the $1$ user setting $R = 0.08$ and $\Lambda \in \myset{3, 10, 100}$ and in the $3{,}400$ user setting $R=0.04$ with $\Lambda \in \myset{30, 100}$, and trained for $150$ epochs.

For \softmaxname-based experiments, we utilized PSGD with a batch size of $20$ and trained for $150$ epochs.
For CIFAR-10, we chose $R \in \myset{0.1, 0.4, 0.6, 1.0}$ and $\Lambda \in \myset{1, 3, 10, 30}$;
for the variant where we protect the whole local dataset, we chose $\Lambda \in \myset{0.5, 1}$ and $R \in \myset{1, 3}$ instead.
For CIFAR-100, we chose the parameter combination $(R, \Lambda) \in \{\,$(0.01, 100), (0.03, 30), (0.03, 100), (0.1, 10), (0.1, 30), (0.3, 3), (0.3, 10), (1, 1), (1, 3)$\,\}$.
For EMNIST, we chose in the $1$ user setting $R = 6$ and $\Lambda \in \myset{0.03, 0.1, 0.3}$ and in the $3{,}400$ user setting $R=2$ with $\Lambda \in \myset{0.3, 1, 3}$ and $R=6$ with $\Lambda \in \myset{0.03, 0.1}$.

For the experiments of \Cref{tab:noniid}, we reported the results of the following hyperparameters: (CIFAR10, \ppmlname, regular \& non-iid) $R=0.06, \Lambda=100$ for the dataset multiplier $1$x and $R=0.06, \Lambda=10$ for the dataset multiplier $67$x; (CIFAR10, \softmaxname, regular) $R=1.0, \Lambda=1$ for both dataset multipliers $1$x, $67$x; (CIFAR10, \softmaxname, non-iid) $R=0.6, \Lambda=3$ for both dataset multipliers $1$x, $67$x; (CIFAR100, \softmaxname, regular) $R=1.0, \Lambda=3$ for dataset multiplier $1$x and $R=1.0, \Lambda=1$ for both dataset multiplier $67$x; (CIFAR100, \softmaxname, non-iid) $R=1.0, \Lambda=3$ for both dataset multipliers $1$x, $67$x.

For the gradient averaging (GradAvg) experiments, we utilized the \emph{opacus}\footnote{accessible at \url{https://github.com/pytorch/opacus/}, Apache-2.0 license} PyTorch library \citep{opacus}, which implements DP-SGD \citep{abadi2016deep}. For a fair comparison, we halved the noise scale for the privacy accounting to comply with bounded DP as \emph{opacus} currently uses unbounded DP: it now uses a noise scale proportional to $2C$ instead of $C$ with $C$ as the clipping bound. We loosely adapted our hyperparameters to the ones reported by \citet{tramer2021differentially} who evaluated DP-SGD on SimCLR's embeddings for the CIFAR-10 dataset. In detail, the neural network is a single-layer perceptron with a $6\,144$\,d input and has the following configuration: (CIFAR-10) $61{,}450$ trainable parameters on a $10$\,d output, (CIFAR-100) $614{,}500$ trainable parameters on a $100$\,d output, and (EMNIST) $380{,}990$ trainable parameters on a $62$\,d output. The loss function is the categorical cross-entropy on a softmax activation function, and training has been performed with SGD. We set the learning rate to $4$, the Poisson sample rate (CIFAR) $q = \sfrac{1024}{50000}$ (EMNIST) $q = \sfrac{1024}{671585}$  which in expectation samples a batch size of $1024$, trained for $40$ epochs, and norm-clipped the gradients with a clipping bound $C = 0.1$.

In the distributed training scenario, instead of running an end-to-end experiment with full SecAgg users, we evaluate a functionally equivalent abstraction without cryptographic overhead. In our CIFAR experiments, we randomly split the available data points among the users and emulated scenarios where not all data points were needed by taking the first training data points. The validation size remained constant. For GradAvg, we kept a constant expected batch size:
$q' = \sfrac{1024}{20000}$ for $20000$ and $q'' = \sfrac{1024}{5000}$ for $5000$ available data points ($wn$). For GradAvg, we emulated a larger number of users by dividing the noise scale $\sigma$ by $\sqrt{w}$ to the benefit of GradAvg.
Here, the model performance is not expected to differ as the mean of the gradients of one user is the same as the mean of gradients from different users: SGD computes, just as GradAvg, the mean of the gradients. Yet, the noise will increase by a factor of $\sqrt{w}$.

\section{Extended Ablation Study (Centralized Setting)}
\label{app:extended_ablation}

\begin{figure}[!ht]
    \centering
    \subfigure[$\delta = 10^{-5}$]{{\resizebox{\columnwidth}{!}{\input{plots/baseline.pgf}\unskip}}}
    \subfigure[$\delta = 2\cdot 10^{-8} \ll \sfrac{1}{\mathrm{data\_size}}$]{{\resizebox{\columnwidth}{!}{\input{plots/baseline_2e8.pgf}\unskip}}}
    
    \caption{
        CIFAR-10 Accuracy vs. $\eps$ budget of \ppmlname (cf. \Cref{ex:dpsvm}), \softmaxname (cf. \Cref{alg:dpsoftmax}), SVM-SMO where only the optima are perturbed, DP-SGD (single-layer, same as GradAvg+SecSum) \citep{abadi2016deep}, and AMP (SVM with objective perturbation) \citep{iyengar2019towards}. For comparison, we report a non-private SVM baseline.
    }
	\label{fig:baseline}
\end{figure}

\subsection{Setup of the Ablation Study}
For SVM-SMO-based experiments, we used the \emph{liblinear} \citep{fan2008liblinear} library via the Scikit-Learn method \emph{LinearSVC}\footnote{\url{https://scikit-learn.org/stable/modules/generated/sklearn.svm.LinearSVC.html}, BSD-3-Clause license} for classification. \emph{Liblinear} is a fast C++ implementation that uses the SVM-agnostic sequential minimal optimization (SMO) procedure. However, it does not offer a guaranteed and private convergence bound.

More specifically, we used the L2-regularized hinge loss, an SMO convergence tolerance of $\mathrm{tol} \coloneqq 2\cdot10^{-12}$ with a maximum of $10{,}000$ iterations which were seldom reached, and a logarithmically spaced inverse regularization parameter $C \in \big\{\, {\myset{3, 6}\cdot10^{-8}}$, ${\myset{1,2,3,6}\cdot10^{-7}}$, ${\myset{1,2,3,6}\cdot10^{-6}}$, ${\myset{1,2,3,6}\cdot10^{-5}}$, ${\myset{1,2}\cdot10^{-4}} \,\big\}$. To better fit with the \emph{LinearSVC} implementation, the original loss function is rescaled by $\sfrac{1}{\Lambda}$ and $C$ is set to $\sfrac{1}{\Lambda\cdot n}$ with $n$ as the number of data points. Furthermore, for distributed SVM-SMO training, we extended the range of the hyperparameter $C$---whenever appropriate---up to $3\cdot 10^{-3}$ which becomes relevant in a scenario with many users and few data points per user. Similar to \ppmlname-based experiments, the best-performing regularization parameter $C$ was selected for each parameter combination.

The non-private reference baseline uses a linear SVM optimized via SMO with the hinge loss and an inverse regularization parameter $C = 2$ (best performing of $C \in \myset{\le 5\cdot 10^{-5}, 0.5, 1, 2}$).

For the ablation study, we also included the Approximate Minima Perturbation (AMP) algorithm\footnote{reference implementation by the authors: \url{https://github.com/sunblaze-ucb/dpml-benchmark}, MIT license} \citep{iyengar2019towards} which resembles an instance of objective perturbation. There, we used a ($80$--$20$)-train-test split with 10 repeats and the following hyperparameters: $L \in \myset{0.1, 1.0, 34.854}$, eps\_frac $\in \myset{.9, .95, .98, .99}$, eps\_out\_frac $\in \myset{.001, .01, .1, .5}$. We selected ($L = 1$, $eps\_out\_frac$ = $0.001$, $eps\_frac = 0.99$) as a good performing parameter combination for AMP. For better performance, we resembled the GPU-capable \emph{bfgs\_minimize} from the Tensorflow Probability package. To provide better privacy guarantees, we leveraged the results of \citet{kairouz2015composition, murtagh2016complexity} for tighter composition bounds on arbitrary DP mechanisms.

\subsection{Results of the Ablation Study}

For the extended ablation study, we considered the centralized setting (only 1 user) and compared different algorithms and different values for the privacy parameter $\delta$. The results are depicted in \Cref{fig:baseline} and display five algorithms: firstly, the differentially private Support Vector Machine with SGD-based training \ppmlname (cf.~\Cref{ex:dpsvm} within BlindAvg of \Cref{alg:dist_blindavg} in the 1 user setting), secondly, the differentially private Softmax-activated single-layer perceptron with SGD-based training \softmaxname (cf.~\Cref{alg:dpsoftmax} within BlindAvg of \Cref{alg:dist_blindavg} in the 1 user setting), thirdly, a similar differentially private SVM but with SMO-based training which does not offer a guaranteed and private convergence bound, fourthly, differentially private Stochastic Gradient descent (DP-SGD) \citep{abadi2016deep} applied on a 1-layer perceptron with the cross-entropy loss, and fifthly, approximate minima perturbation (AMP) \citep{iyengar2019towards} which is based upon an SVM with objective perturbation. Note that only SVM-SMO, \ppmlname, and \softmaxname have an output sensitivity and are thus suited for this efficient \blindavg scheme.

While all algorithms come close to the non-private baseline with rising privacy budgets $\eps$, we observe that although DP-SGD performs best, SVM-SMO and \softmaxname come considerably close, \ppmlname has a disadvantage above SVM-SMO of about a factor of $2$, and AMP a disadvantage of about a factor of $4$. We suspect that DP-SGD is able to outperform the variants other than \softmaxname as it directly optimizes for the multi-class objective via the cross-entropy loss, while others are only able to simulate it via the one-vs-rest (ovr) SVM training scheme. Additionally, DP-SGD has a noise-correcting property from its iterative noise application. The inherently multi-class \softmaxname performs better than ovr-based \ppmlname which indicates that a joint learning of all classes can boost performance. \softmaxname additionally has a privacy advantage as it does not need to rely on sequential composition as it has an output sensitivity for all classes which is another factor that can lead to the boost of \softmaxname above \ppmlname. Although SVM-SMO also has an output sensitivity and renders better than \ppmlname, it does not offer a privacy guarantee when convergence is not reached. In the case of AMP, we have an inherent disadvantage of about a factor of $3$ due to an unknown output distribution, and thus bad composition results in the multi-class SVM. Here, the privacy budget of AMP roughly scales linearly with the number of classes.

For DP-SGD, \ppmlname, \softmaxname, and SVM-SMO, \Cref{fig:baseline} shows a considerable but limited effect on the privacy budget $\eps$ for a smaller and considerably more secure privacy parameter $\delta \ll \sfrac{1}{nw}$, where $nw$ is the sum of the size of all local datasets.

\section{Proofs}
\label{app:proofs}

\subsection{Security Guarantee of \blindavg}

\subsubsection{Proof of \Cref{cor:ampl_avg} (Privacy Amplification via Averaging)}
\label{sec:proof_ampl_avg}

\begin{replemma}{cor:ampl_avg}
    With the notation in \Cref{def:conf}, \blindavg{} of \Cref{alg:dist_blindavg} without noise, $\avg(n^{(i)} \cdot T_\xi(D^{(i)}))$, has a sensitivity of $s' \cdot \sfrac{1}{w}$ for each model if $s = \sfrac{s'}{n}$.

\end{replemma}

\begin{proof}
Without loss of generality, we consider one arbitrary model which corresponds to one class $k \in \myset{1,\dots,K}$ for \ppmlname and all classes $k \coloneqq K$ for \softmaxname. We know that $T_\xi$ is an $s$-sensitivity bounded algorithm thus
\begin{align}
    s = \max_{D^{(i)}_0 \sim D^{(i)}_1}\max_r \Abs{T_\xi(D^{(i)}_0,r) - T_\xi(D^{(i)}_1,r)}
\end{align}
with $D^{(i)}_0$ and $D^{(i)}_1$ as $1$-neighboring datasets and $r$ as the randomness of $T$ (cf. \Cref{thm:m2} for details about a randomized sensitivity involving $r$). For instance, for $T = \text{\ppmlname}$ we have $s = \frac{2(\Lambda R+c)}{\Lambda n^{(i)}}$ (cf.~\Cref{ex:dpsvm}) and for $T = \text{\softmaxname}$ we have $s = \frac{2(\Lambda R+\sqrt{2}c)}{\Lambda n^{(i)}}$ (cf. \Cref{thm:sens_softmax}) which fulfill the condition $s \propto \sfrac{s'}{n^{(i)}}$.

By \Cref{alg:dist_blindavg}, we take the average of multiple local models, i.e. $\avg(n^{(i)} \cdot T_\xi(D^{(i)})) = \frac{1}{w} \sum_{i=1}^{w} n^{(i)} \cdot T_\xi(D^{(i)}, r)$. The challenge element---i.e., the  element that differs between $D^{(i)}_0$ and $D^{(i)}_1$---is only contained in one of the $w$ models. By the application of the parallel composition theorem, we know that the sensitivity reduces to 
\begin{align}
    &\nonumber\phantom{=~} \max_{\forall i=0,\dots,w \colon D^{(i)}_0 \sim D^{(i)}_1}\max_r \Bigg\lvert\frac{1}{w}\sum_{i=1}^{w} n^{(i)} \cdot T_\xi(D^{(i)}_0,r) \\
    &\nonumber\phantom{= \max_{\forall i=0,\dots,w \colon D^{(i)}_0 \sim D^{(i)}_1}\max_r \Bigg\lvert~} - \frac{1}{w}\sum_{i=1}^{w} n^{(i)} \cdot T_\xi(D^{(i)}_1,r)\Bigg\rvert \\
    &\nonumber= \max_{\forall i=0,\dots,w \colon D^{(i)}_0 \sim D^{(i)}_1}\max_r \Abs{\frac{n^{(i)}}{w} ( T_\xi(D^{(i)}_0,r) - T_\xi(D^{(i)}_1,r))} \\
    &\le \max_{\forall i=0,\dots,w \colon D^{(i)}_0 \sim D^{(i)}_1}\max_r \left( \frac{n^{(i)}}{w} s \right) = s' \cdot \frac{1}{w}.
\end{align}
Hence, the constant $\sfrac{n^{(i)}}{w}$ factor reduces the sensitivity by a factor of $\sfrac{n^{(i)}}{w}$.
\end{proof}

\subsubsection{Proof of \Cref{thm:sqrtgauss} (Distribute Noise to Users)}
\label{sec:proof_sqrtgauss}

We recall \Cref{thm:sqrtgauss}:

\begin{replemma}{thm:sqrtgauss}
    With the notation in \Cref{def:conf} and noise scale $\Tilde{\sigma}$:
$\frac{1}{w}\sum_{i=1}^{w} \mathcal{N}(0,(\Tilde{\sigma}\cdot\sfrac{1}{\sqrt{w}})^2) = \mathcal{N}(0,(\Tilde{\sigma}\cdot\sfrac{1}{w})^2)$.

\end{replemma}

\begin{proof}
We have to show that
\begin{align*}
    \textstyle \frac{1}{w}\sum_{i=1}^{w} \mathcal{N}(0,(\Tilde{\sigma} \cdot \frac{1}{\sqrt{w}})^2) = \mathcal{N}(0,(\Tilde{\sigma} \cdot \frac{1}{w})^2).
\end{align*}
It can be shown that the sum of normally distributed random variables behaves as follows: Let $X \sim \mathcal{N}(\mu_X, \sigma^2_X)$ and $Y \sim \mathcal{N}(\mu_Y, \sigma^2_Y)$ two independent normally-distributed random variables, then their sum $Z = X + Y$ equals $Z \sim \mathcal{N}(\mu_X + \mu_Y, \sigma_X^2 + \sigma_Y^2)$ in the expectation.

Thus, in this case, we have
\begin{align}
    \nonumber\textstyle \frac{1}{w}\sum_{i=1}^{w} \mathcal{N}(0,(\Tilde{\sigma} \cdot \frac{1}{\sqrt{w}})^2) 
    &\textstyle= \frac{1}{w}\mathcal{N}(0, w\cdot(\Tilde{\sigma} \cdot \sfrac{1}{\sqrt{w}})^2) \\
    &\textstyle= \frac{1}{w}\mathcal{N}(0, \Tilde{\sigma}^2).
\end{align}
As the normal distribution belongs to the location-scale family, we get $\mathcal{N}(0,(\Tilde{\sigma} \cdot \sfrac{1}{w})^2)$.
\end{proof}

\subsubsection{Proof of \Cref{thm:distributed_blindavg} (\blindavg is computational-DP)}
\label{sec:blindavg_is_dp}

We state the full version of \Cref{thm:distributed_blindavg}:

\begin{reptheorem}{thm:distributed_blindavg}
    With the notation in \Cref{def:conf}, a maximum fraction of dropouts $\rho \in [0,1]$, and a maximum fraction of corrupted users $\gamma \in [0,1]$. Assume that secure summation $\pi_{\mathrm{SecSum}}$ exists as in \Cref{def:secagg}.
Then \blindavg{} (cf.~\Cref{alg:dist_blindavg}) satisfies computational $(\eps, \delta+\nu_1)$-DP for all neighboring central datasets $\mho,\mho'$ with $\delta(\eps, K_{\mathrm{comp}})$ as in \Cref{def:deltagauss}, for $\nu_1 \coloneqq (1 + \exp(\eps))\cdot\nu(\eta)$ and a function $\nu$ negligible in the security parameter $\eta$ used in $\pi_{\mathrm{SecSum}}$.

\end{reptheorem}

\begin{proof}
We first show $(\varepsilon,\delta)$-DP for a variant $M_1$ of \blindavg that uses the ideal summation protocol $\mathcal{F}$ instead of $\pi_{\mathrm{SecSum}}$. We conclude that for \blindavg (abbreviated as $M_2$) which uses the real secure summation protocol $\pi_{\mathrm{SecSum}}$ for some negligible function $\nu_1$ $(\varepsilon,\delta + \nu_1)$-DP holds.

Recall that we assume at least $\noncolluding\cdot w$ many honest users. As we solely rely on the honest $\noncolluding\cdot w$ to contribute correctly distributed noise to the learner $T$, we have for each output model similar to \Cref{thm:sqrtgauss}
\begin{align}
    &\nonumber\frac{1}{w}\sum_{i=1}^{\noncolluding\cdot w} \mathcal{N}(0,(\Tilde{\sigma} \cdot \frac{1}{\sqrt{w}})^2)
    = \sum_{i=1}^{\noncolluding\cdot w} \mathcal{N}(0,(\Tilde{\sigma} \cdot \frac{1}{w\sqrt{w}})^2)\\
    &= \mathcal{N}(0,(\Tilde{\sigma}\cdot \frac{\sqrt{\noncolluding\cdot w}}{w\sqrt{w}})^2)
    = \mathcal{N}(0,(\Tilde{\sigma}\cdot \frac{\sqrt{\noncolluding}}{w})^2).
\end{align}
Hence, we scale the noise parameter $\tilde \sigma$ with $1/\sqrt{\noncolluding}$ and get
\begin{align}
    \frac{1}{w}\sum_{i=1}^{\noncolluding w} \mathcal{N}(0,(\Tilde{\sigma} \cdot \frac{1}{\sqrt{t}} \cdot \frac{1}{\sqrt{w}})^2)
    = \mathcal{N}(0,(\Tilde{\sigma} \cdot  \frac{1}{w})^2).
\end{align}

By \Cref{cor:ampl_avg}, \Cref{thm:sqrtgauss}, and \Cref{lem:gauss_mech}, we know that $M_1$ satisfies $(\eps,\delta)$-DP (with the parameters as described above).

Considering an unbounded attacker $\mathcal{A}$, we know that for any pair of neighboring data sets $\mho,\mho'$ the following holds
\[
    \Pr\left[\mathcal{A}\left(\mathcal{M}_1(\mho)\right) = 1\right] \le \exp(\varepsilon) \Pr\left[\mathcal{A}\left(\mathcal{M}_1(\mho')\right) = 1\right] + \delta
\]

If $\pi_{\mathrm{SecSum}}$ is a secure summation protocol, then there is a negligible function $\nu$ such that the following holds w.l.o.g. for any neighboring data sets $\mho, \mho'$ (differing in at most one element):
\begin{align}
    \Pr\left[\mathcal{A}\left(\mathcal{M}_2(\mho)\right) = 1\right] - \nu(\eta) 
    \le \Pr\left[\mathcal{A}\left(Sim_{\mathcal{F}}(\mathcal{M}_1(\mho))\right) = 1\right].
\end{align}
For the attacker $\mathcal{A}'$ that first applies $\mathrm{Sim}$ and then $\mathcal{A}$, we get:
\begin{align}
    &\nonumber\phantom{\le~} \Pr\left[\mathcal{A}\left(\mathcal{M}_2(\mho)\right) = 1\right] - \nu(\eta) \\
    &\nonumber\le \exp(\varepsilon)\Pr\left[\mathcal{A}\left(Sim_{\mathcal{F}}(\mathcal{M}_1(\mho'))\right) = 1\right] + \delta \\
    &\le \exp(\varepsilon) \left(\Pr\left[\mathcal{A}\left(\mathcal{M}_2(\mho')\right) = 1\right] + \nu(\eta)\right)  + \delta
    \intertext{thus we have}
    &\nonumber\phantom{\le~} \Pr\left[\mathcal{A}\left(\mathcal{M}_2(\mho)\right) = 1\right] \\
    &\le \exp(\varepsilon) \Pr\left[\mathcal{A}\left(\mathcal{M}_2(\mho')\right) = 1\right] + \delta + \left(1 + \exp(\varepsilon)\right) \cdot \nu(\eta).
    \intertext{From a similar argumentation, it follows that}
    &\nonumber\phantom{\le~} \Pr\left[\mathcal{A}\left(\mathcal{M}_2(\mho')\right) = 1\right] \\
    &\le \exp(\varepsilon) \Pr\left[\mathcal{A}\left(\mathcal{M}_2(\mho)\right) = 1\right] + \delta + (1 + \exp(\varepsilon)) \cdot \nu(\eta)
\end{align}
holds.

Hence, with $\nu_1 \coloneqq (1 + \exp(\varepsilon)) \cdot \nu(\eta)$ the mechanism \blindavg mechanism $\mathcal{M}_2$ which uses $\pi_{\mathrm{SecSum}}$ is $(\varepsilon, \delta + \nu_1)$-DP. As $\nu$ is negligible and $\eps$ is constant, $\nu_1$ is negligible as well.
\end{proof}

\begin{corollary}
\label{cor:blindavg_dp_secagg}
    With the notation in \Cref{def:conf}, a maximum fraction of dropouts $\rho \in [0,1]$, and a maximum fraction of corrupted users $\gamma \in [0,1]$, if secure authentication encryption schemes and authenticated key agreement protocol exist, then \blindavg{} (cf.~\Cref{alg:dist_blindavg}) instantiated with $\pi_{\mathrm{SecSum}} = \pi_{\mathrm{SecAgg}}$ \citep{bell2020secure} satisfies computational $(\eps, \delta+\nu_1)$-DP with $\delta(\eps, K_{\mathrm{comp}})$ as in \Cref{def:deltagauss}, for $\nu_1 \coloneqq (1 + \exp(\eps))\cdot\nu(\eta)$ and a function $\nu$ negligible in the security parameter $\eta$ used in secure summation.

\end{corollary}

This follows directly from \Cref{thm:distributed_blindavg}, as by \Cref{theorem:secure_aggregation}, we know that $\pi_{\mathrm{SecAgg}}(s_1, \dots, s_n)$ securely emulates $\mathcal{F}$ (w.r.t. an unbounded attacker).

\subsubsection{Proof of \Cref{cor:userlevel_sens} (User-level Sensitivity)}
\label{sec:userlevel_sens}

We recall \Cref{cor:userlevel_sens}:

\begin{repcorollary}{cor:userlevel_sens}
    With the notation in \Cref{def:conf}, we say that learner $T_\xi$ is $R$-norm bounded if for any local dataset $D^{(i)}$: $\norm{T_\xi(D^{(i)})} \le R$. Any $R$-norm bounded $T_\xi$ has a deterministic sensitivity $s = 2R$.
Then, $D^{(i)} \mapsto T_\xi(D^{(i)})\allowbreak + \mathcal{N}(0,\Tilde{\sigma}^2 s^2 I_{p \times K})$ satisfies computational $(\Upsilon\eps,\delta)$, $\Upsilon$-group DP for all $\Upsilon$-neighboring central datasets $\mho,\mho'$ if $\forall i,j\colon n^{(i)} = n^{(j)}$ with $\delta(\eps, K_{\mathrm{comp}})$ as in \Cref{def:deltagauss}, $\Upsilon = n^{(i)} = \abs{D^{(i)}}$, and $\nu$ negligible in the security parameter used in $\pi_{\mathrm{SecSum}}$.

\end{repcorollary}

\begin{proof}
We know that the deterministic sensitivity of learner $T_\xi$ is defined as $s = \max\limits_{D\sim D'}\lVert T_\xi(D^{(i)}) - T_\xi(D'^{(i)}) \rVert$ for $\Upsilon$-neighboring datasets $D^{(i)},D'^{(i)}$ of the $i$-th user. Thus, in our case we have $s = 2R$ since for any dataset $\Tilde{D}$, we have $T_\xi(\Tilde{D}) \in [-R,R]$. As this holds independent on the dataset $\Tilde{D}$ and by \Cref{lem:gauss_mech} and by \Cref{lem:group_privacy},  we can protect any arbitrary number of data points per user, i.e., we have $\Upsilon$-group DP.
\end{proof}

\subsection{Non-interactive Blind Model Averaging (\blindavg)}

\subsubsection{Proof of \Cref{lem:averagerepresenter} (Averaged Representer Theorem)}
\label{app:proof_averagerepresenter}

We recall \Cref{lem:averagerepresenter}:

\begin{repcorollary}{lem:averagerepresenter}
    If a local learner $T_\xi$ on dataset $D^{(i)}$ admits a solution of the form $f^{(i)} = T_\xi(D^{(i)}) = \sum_{j=1}^{n} \alpha_j^{(i)} x_j^{(i)}$ (cf. \Cref{thm:representer}) then the average $f \coloneqq \avg(f^{(i)})$ admits a solution of the form $f = T_\xi(\mho) = \frac{1}{w}\sum_{i=1}^{w}\sum_{j=1}^{n} \alpha_{j}^{(i)} x_{j}^{(i)}$ with $\mho \coloneqq \bigcup_{i=1}^w D^{(i)}$.

\end{repcorollary}

\begin{proof}
\begin{align}
    \nonumber\textstyle \avg(f^{(i)}) &= \textstyle\frac{1}{w}\sum_{i=1}^{w} T_\xi(D^{(i)}) \\
    &\nonumber= \textstyle\frac{1}{w}\sum_{i=1}^{w}\sum_{j=1}^{N} \alpha_{j}^{(i)} x_{j}^{(i)} \\
    &= \textstyle T_\xi(\mho) = f.
\end{align}
\end{proof}

\subsubsection{Proof of \Cref{lem:supportvec} (Support Vectors of Averaged SVMs)}
\label{app:proof_supportvec}

We recall \Cref{lem:supportvec}:

\begin{replemma}{lem:supportvec}
    A locally trained hinge-loss linear SVM has the support vectors $V^{(i)} \subseteq D^{(i)}$. Then, the average $\frac{1}{w}\sum_{i=1}^w f^{(i)}$ has the support vectors $V = \bigcup_{i=1}^{w} V^{(i)}$.

\end{replemma}

\begin{proof}
A learning problem that is based on a hinge-loss SVM fulfills the representer theorem requirements due to the L2-regularized ERM objective function. In fact, if a data point $x_j$ is a support vector, i.e. $x_j \in V$, then after successful training its corresponding $\alpha_j$ is restricted by $0 < \alpha_j \le \Lambda \land y_j=1$ or $0 > \alpha_j \ge -\Lambda \land y_j=-1$, or $\alpha_j=0$ \citep[Equation 28-30]{CS229lecturenotes}. Thus, we denote $V^{(i)} = \myset{x_j^{(i)} \in D^{(i)} \mid \alpha_j^{(i)} \neq 0}$. By \Cref{lem:averagerepresenter}, we have that the average of locally trained models $\avg(T_\xi(D^{(i)})) = \frac{1}{w}\sum_{i=1}^{w}\sum_{j=1}^{n} \alpha_{j}^{(i)} x_{j}^{(i)}$.
Since the local datasets are disjoint we simplify $\frac{1}{w}\sum_{i=1}^{w}\sum_{j=1}^{n} \alpha_{j}^{(i)} x_{j}^{(i)} = \frac{1}{w}\sum_{j=1}^{\abs{\mho}} \alpha_{j} x_{j}$ for the combined local datasets $\mho = \bigcup_{i=1}^{w} D^{(i)}$ and a flattened $\alpha = \begin{bmatrix} \alpha_1^{(1)} & \dots & \alpha_n^{(1)} & \alpha_1^{(2)} & \dots & \alpha_n^{(w)} \end{bmatrix}$.
A model which is represented by $\frac{1}{w}\sum_{j=1}^{\abs{\mho}} \alpha_{j} x_{j}$ has the support vectors $V = \myset{x_j \in \mho \mid \alpha_j \neq 0} = \bigcup_{i=1}^{w} V^{(i)}$, as the support vector characteristic is uniquely determined by $\alpha$ and each local $\alpha_j^{(i)}$ is element of $\alpha$ and responsible for the same data point.
\end{proof}

\subsubsection{Proof of \Cref{cor:convergenceOfBlindlyAveraging} (Averaging Locally Trained SVMs Converges)}
\label{app:proof_avg_converges}

We recall \Cref{cor:convergenceOfBlindlyAveraging}:

\begin{reptheorem}{cor:convergenceOfBlindlyAveraging}
    Given the same local data sizes $\forall{i,i'}\colon n^{(i)} = n^{(i')}$, there exists a regularization parameter $\Lambda$ such that the average of HingeSVMs trained with projected subgradient descent using weighted averaging $T=\text{HingeSVM-SGDWA}$ converges with the number of local iterations $M$ to the best central model with $\mho \coloneqq \bigcup_{i=1}^w D^{(i)}$, i.e.
$\Eop[\mathcal{J}(\avg(\text{HingeSVM-SGDWA}), \mho) - \inf_{f} \mathcal{J}(f,\mho)] \in \mathcal{O}(\sfrac{1}{M})$.

\end{reptheorem}

\begin{proof}
First (1), we show that there exists a regularization parameter $\Lambda$ for which the converged global model equals the average of the converged locally trained models: $T_\xi(\mho) = \avg(T\_xi(D^{(i)}))$. Second (2), we show that both the global and the local models converge with rate $\mathcal{O}(\sfrac{1}{M})$.

Note that we assume that each data point $x_j$ is structured as $[1, x_{j,1}, \dots, x_{j,p}]$ to include the intercept. We also denote the flattened $\alpha^{(\mathrm{avg\_loc})} = \begin{bmatrix} \alpha_1^{(1)} & \dots & \alpha_n^{(1)} & \alpha_1^{(2)} & \dots & \alpha_n^{(w)} \end{bmatrix}$ as the dual coefficients of the averaged local SVM and $\alpha^{(\mathrm{glob})}$ as the dual coefficients of the global SVM.
 	
(1) By \Cref{lem:supportvec} we know for the combined local datasets $\mho = \bigcup_{i=1}^{w} D^{(i)}$ that
\begin{align}
	\avg(T_\xi(D^{(i)})) = \frac{1}{wN}\sum_{j=1}^{\abs{\mho}} \alpha^{(\mathrm{avg\_loc})}_{j} x_{j} = \frac{1}{\abs{\mho}}\sum_{j=1}^{\abs{\mho}} \alpha^{(\mathrm{avg\_loc})}_{j} x_{j}.
\end{align}
Note that we assume a scaled parameter per local SVM: $T_\xi(D^{(i)}) = \frac{1}{n} \sum_{j=1}^n \alpha_j x_j$. Without this assumption, we would not average the local SVMs but instead compute their sum.

For the global model, we write by the representer theorem
\begin{align}
	T(\mho) = \frac{1}{\abs{\mho}}\sum_{j=1}^{\abs{\mho}} \alpha^{(\mathrm{glob})}_{j} x_{j}.
\end{align}
Thus, by parameter comparison we have that $T_\xi(\mho) = \avg(T_\xi(D^{(i)}))$ if $\forall j\colon \alpha^{(\mathrm{glob})}_{j} = \alpha^{(\mathrm{avg\_loc})}_{j}$. By the characteristic of a hinge-loss linear SVM, we know that any $\alpha_j$ has the value $\alpha_j = \Lambda y_j$ if a data point is a support vector inside the margin \citep[Equation 28-30]{CS229lecturenotes}. Hence, $\forall j\colon \alpha^{(\mathrm{glob})}_{j} = \alpha^{(\mathrm{avg\_loc})}_{j}$ if the margin is large enough that for both SVMs all data points are inside the margin. Since the margin of a hinge-loss linear SVM is the inverse of the parameter norm, $\norm{T(\mho)}^{-1}$, and the parameter norm gets smaller with an increased regularization parameter $\Lambda$ by the definition of the objective function $\frac{1}{n} \sum_{(x,y) \in D^{(i)}} \max(0, 1- y \inner{f}{x}) + \Lambda\inner{f}{f}$, we derive that there exists a regularization parameter $\Lambda$ which is large enough s.t. all data points are within the margin.
	
(2) By \citet{lacoste2012simpler}, we know that a hinge-loss linear SVM converges to the optima with rate $\mathcal{O}(M^{-1})$, if we use projected subgradient descent using weighted averaging (SGDWA) as an optimization algorithm, i.e.
$\E{\mathcal{J}(\avg(\text{HingeSVM-SGDWA}(D^{(i)})), \mho) - \inf_{f} \mathcal{J}(f,\mho)} \in \mathcal{O}(\sfrac{1}{M})$.
\end{proof}

\subsection{Additional Privacy Proofs}

\subsubsection{Proof of \Cref{thm:m2} (Learner $T$ with a Randomized Sensitivity is DP)}
\label{app:randomized_sens}

\begin{lemma}
\label{lem:m1}
    Let $T_{\mathrm{priv}}: (D,r,\kappa) \to U$ be a randomized mechanism on dataset $D$ with two independent randomnesses $r$ and $\kappa$ and universe $U$. We define $T^r_{\mathrm{priv}}(D, \kappa) \coloneqq T_{\mathrm{priv}}(D,r, \kappa)$, i.e., $T^r_{\mathrm{priv}}: (D, \kappa) \to U$. If $\forall r$, $T^r_{\mathrm{priv}}$ is $(\eps,\delta$)-DP, then $T_{\mathrm{priv}}$ is $(\eps,\delta$)-DP.

\end{lemma}

\begin{proof}
Let $D, D'$ be neighboring datasets and $S \subseteq U$ be defined over some universe $U$ as required. Let $R$ denote a distribution of randomness $r$ which is independent of the data $D$ as $r$ and $D$ are separate inputs.
We show that if $\forall r\colon \Pr\limits_{\kappa}[T^r_{\mathrm{priv}}(D,\kappa) \in S ] \le \exp(\eps) \Pr\limits_{\kappa}[T^r_{\mathrm{priv}}(D',\kappa) \in S ] + \delta$ then $\Pr\limits_{r,\kappa}[T_{\mathrm{priv}}(D,r,\kappa) \in S ] \le \exp(\eps)  \Pr\limits_{r,\kappa}[T_{\mathrm{priv}}(D',r,\kappa) \in S ] + \delta$. The proof is similar to that of \citet[Lemma 5]{wu2017bolt}.

By the law of total probability, we have
\begin{align}
    &\nonumber\phantom{=~} \Pr_{r,\kappa}[T_{\mathrm{priv}}(D,r,\kappa) \in S ] \\
    &\nonumber= \sum_r \Pr[R = r] \Pr_\kappa[T_{\mathrm{priv}}(D,r, \kappa) \in S \mid R = r] \\
    &\nonumber= \sum_r \Pr[R = r] \Pr_\kappa[T^r_{\mathrm{priv}}(D,\kappa) \in S] \\
    &\nonumber\le \sum_r \Pr[R = r] \left(\exp(\eps) \Pr_\kappa[T^r_{\mathrm{priv}}(D', \kappa) \in S] + \delta\right) \\
    &\nonumber= \exp(\eps) \sum_r \Pr[R = r] \Pr_\kappa[T_{\mathrm{priv}}(D', r, \kappa) \in S \mid R = r] \\
    &\nonumber\qquad\qquad + \sum_r \Pr[R = r] \delta \\
    &= \exp(\eps) \Pr_{r,\kappa}[T_{\mathrm{priv}}(D', r, \kappa) \in S] + \delta.
\end{align}
\end{proof}

\begin{theorem}
\label{thm:m2}
    Let $T_{\mathrm{priv}}: (D,r) \mapsto T(D,r) + \kappa$ be an additive mechanism with a Gaussian randomness $\kappa \in \pdf_{\mathcal{N}(0,\sigma^2)}$ and noise scale $\sigma$ where $T$ is a randomized mechanism with randomness $r$ and dataset $D$. $T$ has a randomized sensitivity $\max_{D, D'}\max_{r} \norm{T(D,r) - T(D',r)} \le s$ where $D, D'$ are $1$-neighboring datasets. Then $T_{\mathrm{priv}}$ is $(\eps,\delta)$-DP.

\end{theorem}

\begin{proof}
Let $R$ denote the distribution of randomness $r$ which by construction does not depend on data $D$ or randomness $\kappa$. We define $T^r(D) \coloneqq T(D,r)$, i.e., $T^r: D \mapsto T(D,r)$. We make a case distinction over each $r \in R$:

For each $r \in R$, we have the mechanism  $T^r_{\mathrm{priv}}: D \mapsto T^r(D) + \kappa$ with a deterministic sensitivity $\max_{D, D'}\max_{r \in R} \norm{T(D,r) - T(D',r)} = \max_{D, D'} \norm{T^r(D) - T^r(D')}$, where $D, D'$ are $1$-neighboring datasets. By construction, $T^r_{\mathrm{priv}}$ is a Gaussian mechanism which is $(\eps,\delta)$-DP by \Cref{lem:gauss_mech}.
By \cref{lem:m1}, since $T^r_{\mathrm{priv}}$ is $(\eps,\delta)$-DP for all $r$, $T_{\mathrm{priv}}$ is $(\eps,\delta)$-DP.
\end{proof}

The same holds if we use the Gaussian mechanism in the group privacy extension (cf. \Cref{lem:group_privacy}) or in the distributed setting (cf. \Cref{cor:ampl_avg}. In each case, we divide the algorithm output by a constant factor $\mathrm{const}$ which scales both the deterministic and the randomized sensitivity by $\mathrm{const}$:
\begin{align}
    &\nonumber\max_{D, D'}\max_{r \in R} \norm{\frac{T(D,r)}{\mathrm{const}} - \frac{T(D',r)}{\mathrm{const}}} \\
    &\nonumber\qquad= \frac{1}{\mathrm{const}}\max_{D, D'}\max_{r \in R} \norm{T(D,r) - T(D',r)} \\
    &\qquad= \frac{s_{\mathrm{rand}}}{\mathrm{const}} \\
    &\nonumber\max_{D, D'} \norm{\frac{T^r(D)}{\mathrm{const}} - \frac{T^r(D')}{\mathrm{const}}} \\
    &\nonumber\qquad= \frac{1}{\mathrm{const}}\max_{D, D'} \norm{T^r(D) - T^r(D')} \\
    &\qquad= \frac{s_{\mathrm{det}}}{\mathrm{const}}.
\end{align}

\subsubsection{Proof of \Cref{lem:group_privacy} (Group Privacy Reduction of a Multivariate Gaussian)}
\label{sec:group_privacy}

\begin{lemma}
\label{lem:group_privacy}
    Let $\pdf_{\mathcal{N}(A, B)}[x]$ denote the probability density function of the multivariate Gaussian distribution with location and scale parameters $A, B$ which is evaluated on an atomic event $x$. For any atomic event $x$, any covariance matrix $\Sigma$, any group size $k \in \mathbb{N}$, and any mean $\mu$, we get
\begin{align*}
    \frac{\pdf_{\mathcal{N}(0, k^2 \Sigma)}[x]}{\pdf_{\mathcal{N}(\mu, k^2 \Sigma)}[x]}
    = \frac{\pdf_{\mathcal{N}(0, \Sigma)}[\sfrac{x}{k}]}{\pdf_{\mathcal{N}(\sfrac{\mu}{k}, \Sigma)}[\sfrac{x}{k}]}.
\end{align*}

\end{lemma}

\begin{proof}
\begin{align}
    &\nonumber\phantom{=~} \frac{\pdf_{\mathcal{N}(0, k^2 \Sigma)}[x]}{\pdf_{\mathcal{N}(\mu, k^2 \Sigma)}[x]} \\
    &\nonumber= \frac{
        \frac{1}{\det(2\pi k^2\Sigma)} \exp(-\frac{1}2 x^T k^2 \Sigma^{-1} x)
    }{
        \frac{1}{\det(2\pi k^2\Sigma)} \exp(-\frac{1}2 \underbrace{(x - \mu)^T k^2 \Sigma^{-1} (x - \mu)}_{\substack{
                = x^T k^2 \Sigma^{-1} x - \mu^T k^2 \Sigma^{-1} x \\ \qquad - x^T k^2 \Sigma^{-1} \mu + \mu^T k^2 \Sigma^{-1} \mu
            }})
    }\\
    &\nonumber=  \exp(-\frac{1}2 (- \mu^T k^2 \Sigma^{-1} x - x^T k^2 \Sigma^{-1} \mu + \mu^T k^2 \Sigma^{-1} \mu))\\
    &\nonumber=  \exp(-\frac{1}2 k^2 \cdot (- \mu^T \Sigma^{-1} x - x^T \Sigma^{-1} \mu + \mu^T \Sigma^{-1} \mu))\\
    \intertext{for $\mu_1 \coloneqq \sfrac{\mu}{k}$}
    &\nonumber=  \exp(-\frac{1}2 \cdot k (- \mu_1^T \Sigma^{-1} x - x^T \Sigma^{-1} \mu_1 + \mu_1^T \Sigma^{-1} \mu_1/k))\\
    \intertext{for $x_1 \coloneqq \sfrac{x}{k}$}
    &\nonumber=  \exp(-\frac{1}2 \cdot (- \mu_1^T \Sigma^{-1} x_1 - x_1^T \Sigma^{-1} \mu_1 + \mu_1^T \Sigma^{-1} \mu_1))\\
    &\nonumber=  \exp(-\frac{1}2 \cdot (- \mu_1^T \Sigma^{-1} x_1 - x_1^T \Sigma^{-1} \mu_1 + \mu_1^T \Sigma^{-1} \mu_1))\\
    &\nonumber=  \frac{
        \frac{1}{\det(2\pi \Sigma)} \exp(-\frac{1}2 x_1^T \Sigma^{-1} x_1)
    }{
        \frac{1}{\det(2\pi \Sigma)} \exp(-\frac{1}2 (x_1 - \mu_1)^T k^2 \Sigma^{-1} (x_1 - \mu_1))
    }\\
    &= \frac{\pdf_{\mathcal{N}(0, \Sigma)}[\sfrac{x}{k}]}{\pdf_{\mathcal{N}(\sfrac{\mu}{k}, \Sigma)}[\sfrac{x}{k}]}
\end{align}
\end{proof}

As the Gaussian distribution belongs to the location-scale family, \Cref{lem:group_privacy} directly implies that the $(\eps, \delta)$-DP guarantees of using $\mathcal{N}(0,k^2\ \Sigma)$ noise for sensitivity $k$ and using $\mathcal{N}(0, \Sigma)$ for sensitivity $1$ are the same.

\subsubsection{Proof of \Cref{lem:tight_adp_gauss} (Representing a Multivariate Gaussian as Univariate Ones)}
\label{app:multivariate_gaussian_as_univariate}

For completeness, we rephrase a proof that we first saw in \citet{abadi2016deep} that argues that sometimes the multivariate Gauss mechanism can be reduced to the univariate Gauss mechanism.

\begin{lemma}
\label{lem:tight_adp_gauss}
    Let $\pdf_{\mathcal{N}(\mu,\diag(\sigma^2))}$ denote the probability density function of a multivariate ($p \ge 1$) spherical Gaussian distribution with location and scale parameters $\mu\in\mathbb{R}^p,\sigma\in\mathbb{R}^p_+$. Let $M_{gauss,p,q}$ be the $p$ dimensional Gaussian mechanism $D \mapsto q(D) + \mathcal{N}(0,\sigma^2 \cdot I_{p})$ for $\sigma^2 > 0$ of a function $q: \mathcal{D} \to \mathbb{R}^p$, where $\mathcal{D}$ is the set of datasets. Then, for any $p \ge 1$, if $q$ is $s$-sensitivity-bounded, then for any $p \ge 1$, there is another $s$-sensitivity-bounded function $q': \mathcal{D} \to \mathbb{R}$ such that the following holds: for all $\eps \ge 0, \delta \in [0,1]$ if $M_{gauss, 1,q'}$ satisfies $(\eps,\delta)$-DP, then $M_{gauss,p,q}$ satisfies $(\eps,\delta)$-DP.

\end{lemma}

\begin{proof}
First observe that for any $s$-sensitivity-bounded function $q''$, two adjacent inputs $D, D'$ (differing in one element) with $\lVert q''(D) - q''(D') \rVert_2 = s$ are worst-case inputs. As a spherical Gaussian distribution (covariance matrix $\Sigma = \sigma^2\cdot I_{p\times n}$) is rotation invariant, there is a rotation such that the difference only occurs in one dimension and has length $s$. Hence, it suffices to analyze a univariate Gaussian distribution with sensitivity $s$. Hence, the privacy loss distribution of both mechanisms (for the worst-case inputs) is the same. As a result, for all $\eps \ge 0, \delta \in [0,1]$ (i.e., the privacy profile is the same), if $(\eps,\delta)$-DP holds for the univariate Gaussian mechanism, it also holds for the multivariate Gaussian mechanism.
\end{proof}

\subsection{\softmaxslp}

\subsubsection{Proof of \Cref{thm:strongconvexity_softmax} (Strong Convexity of \softmaxslp)}
\label{app:strongconvexity_softmax}

We state the full version of \Cref{thm:strongconvexity_softmax}:

\begin{reptheorem}{thm:strongconvexity_softmax}
    Let $\mathcal{J}(f,D) \coloneqq \frac\Lambda2 \sum_{k=1}^K \inner{f_k}{f_k} + \frac1n\sum_{(x,y)\in D} \mathcal{L}_{\mathrm{CE}}(y, \inner{f}{x})$ denote the objective function with the cross-entropy loss $\mathcal{L}_{\mathrm{CE}}(y, z) \coloneqq -\sum_{k=1}^K y_k\log\frac{\exp z_k}{\sum_{j=1}^K \exp z_j}$ and parameters $f\in\mathbb{R}^{d+1,K}$, dataset $D$ where $(x,y)\in D$ with data points $x\in\mathbb{R}^{d+1}$ structured as $\begin{bmatrix} 1 & x_1 & \dots & x_d \end{bmatrix}$ and labels $y\in\set{0, 1}^K$, number of classes $K$, and regularization parameter $\Lambda$. $\mathcal{J}$ is $\Lambda$-strongly convex.

\end{reptheorem}

\begin{proof}
$\mathcal{J}$ is $\mu$-strongly convex if $\mathcal{J} - \frac{\mu}{2} \inner{f}{f}$ is convex. In our case, with $\mu = \Lambda$, it remains to be shown show that the cross entropy loss $\mathcal{L}_{\mathrm{CE}}(y,z)$ is convex since a linear layer like $\inner{f}{x}$ represents an affine map which preserves convexity \citep{bertsekas2009convex}.

It is known that the cross entropy loss is convex by a simple argumentation: If the Hessian is positive semidefinite $\nabla^2\mathcal{L}_{\mathrm{CE}}(y,z) \succeq 0$ then $\mathcal{L}_{\mathrm{CE}}$ is convex. By the Gershgorin circle theorem, a symmetric diagonally dominant matrix is positive semi-definite if the diagonals are non-real.

Since the second derivative of the cross-entropy loss is $\frac{\partial^2}{\partial z_{p} \partial z_{q}}\mathcal{L}_{\mathrm{CE}} = s_p(1_{[p=q]} - s_q)$ for the softmax probabilities $s_p = \frac{\exp z_p}{\sum_{j=1}^K \exp z_j}$, we conclude that the diagonals are non-negative since $s_p(1-s_p)$ for $0\le s_p \le 1$ is always non-negative. The Hessian is diagonally dominant if for every row $p$ the absolute value of the diagonal entry is larger than or equal to the sum of the absolute values of all other row entries. In our case, we have
\begin{align}    
    &\nonumber\phantom{\iff~} \forall p\colon \abs{s_p(1-s_q)} \ge \sum_{q=1, q \neq p}^K \abs{s_p(-s_q)} \\
    &\nonumber\iff \forall p\colon (1-s_q) \ge \sum_{q=1, q \neq p}^K s_q \\
    &\iff  \forall p\colon (1-s_q) \ge (1-s_p)
\end{align}
\end{proof}

\subsubsection{Proof of \Cref{thm:lipschitz_softmax} (Lipschitzness of \softmaxslp)}
\label{app:lipschitz_softmax}

We state the full version of \Cref{thm:lipschitz_softmax}:

\begin{reptheorem}{thm:lipschitz_softmax}
    Let $\mathcal{J}(f,D) \coloneqq \frac\Lambda2 \sum_{k=1}^K \inner{f_k}{f_k} + \frac1n\sum_{(x,y)\in D} \mathcal{L}_{\mathrm{CE}}(y, \inner{f}{x})$ denote the objective function with the cross-entropy loss $\mathcal{L}_{\mathrm{CE}}(y, z) \coloneqq -\sum_{k=1}^K y_k\log\frac{\exp z_k}{\sum_{j=1}^K \exp z_j}$ and parameters $f\in\mathbb{R}^{d+1,K}$, dataset $D$ where $(x,y)\in D$ with data points $x\in\mathbb{R}^{d+1}$ structured as $\begin{bmatrix} 1 & x_1 & \dots & x_d \end{bmatrix}$ and labels $y\in\set{0, 1}^K$, number of classes $K$, and regularization parameter $\Lambda$. $\mathcal{J}$ is $L$-Lipschitz with $L=\Lambda R + \sqrt{2}c$ where $\norm{x} \le c$ and $\norm{f} \le R$.

\end{reptheorem}

\begin{proof}
In the following, we abbreviate $d' \coloneqq d+1$, flatten $f \in \mathbb{R}^{d'K}$ and notate $z \coloneqq (x,y)$.

The Lipschitz continuity is defined as
\[
    \textstyle\sup_{z\in D, f,f'} \frac{\norm{\mathcal{J}(f,z) - \mathcal{J}(f',z)}}{\norm{f - f'}} \le L.
\]

We first (1) show
\begin{align}
    \textstyle\sup_{z\in D, f,f'} \frac{\norm{\mathcal{J}(f,z) - \mathcal{J}(f',z)}}{\norm{f - f'}} \le \sup_{z\in D, f}\norm{\nabla_{f}\mathcal{J}(f,z)}
\end{align}
using the mean value theorem and subsequently (2) bound $\sup_{z\in D, f}\norm{\nabla_{f}\mathcal{J}(f,z)} \le L$.

(1) Recall that the multivariate mean value theorem states that for some function $g\colon G \mapsto \mathbb{R}$ on an open subset $G \in \mathbb{R}^n$, some $x,y \in G$ and some $c \in [0,1]$, we have
\begin{align}
    g(y) - g(x) = \inner{\nabla g((1-c)x + cy)}{y-x}.
\end{align}

In our case, we write
\begin{align}
    &\nonumber\textstyle\phantom{=}~\sup_{z\in D, f,f'}\frac{\norm{\mathcal{J}(f,z) - \mathcal{J}(f',z)}}{\norm{f-f'}}
    \intertext{by the multivariate mean value theorem for some $c \in [0,1]$}
    &\nonumber\textstyle= \sup_{z\in D, f,f'}\frac{\Abs{\inner{\nabla \mathcal{J}((1-c)f' - cf,z)}{f-f'}}}{\norm{f-f'}}
    \intertext{for $f'' \coloneqq (1-c)f' - cf$ and by the Cauchy-Schwarz inequality $\abs{\inner{\nabla_{f''} \mathcal{J}(f'', z)}{f-f'}} \le \norm{\nabla_{f''} \mathcal{J}(f'', z)} \cdot \norm{f-f'}$}
    &\textstyle\le \sup_{z\in D, f''}\norm{\nabla_{f''} \mathcal{J}(f'',z)}.
\end{align}

(2) We know that for $1 \le j \le d', 1 \le p \le K$ the partial derivative of $\mathcal{J}$ is $\frac{\partial}{\partial f_p}\mathcal{J}(f,(x,y)) = \Lambda f_{lp} + x_l\cdot(s_p - 1_{[y=p]})$ with $s_p \coloneqq \frac{\exp \inner{f_p}{x}}{\sum_{j=1}^K \exp \inner{f_j}{x}}$.
Thus, we have
\begin{align}
    &\nonumber\phantom{=~}\textstyle\norm{\nabla_f\mathcal{J}(f,z)} \\
    &\nonumber= \sqrt{\sum_{lp=1}^{d'K} \bigl(
        \Lambda f_{lp} + x_l (s_p - 1_{[y=p]})
    \bigr)^2}\\
    &\nonumber= \sqrt{\sum_{lp=1}^{d'K} (
        \Lambda^2 f_{lp}^2 + 2\Lambda f_{lp} x_l (s_p - 1_{[y=p]}) + x_l^2 (s_p - 1_{[y=p]})^2
    )}\\
    &\nonumber= \sqrt{\substack{\displaystyle \Lambda^2\norm{f}^2 +
         2\Lambda \sum_{l=1}^{d'} x_l \sum_{p=1}^K f_{lp} (s_p - 1_{[y=p]}) \\
        \displaystyle\quad + \sum_{l=1}^{d'} x_l^2 \sum_{p=1}^K (s_p - 1_{[y=p]})^2
    }}
    \intertext{due to the Cauchy-Schwarz inequality, we have $\sum_{p=1}^K f_{lp} (s_p - 1_{[y=p]}) \le \sqrt{\sum_{p=1}^K f_{lp}^2} \sqrt{\sum_{p=1}^K (s_p - 1_{[y=p]})^2}$ and $\sum_{l=1}^{d'} x_l \sqrt{\sum_{p=1}^K f_{lp}^2} \le \sqrt{\sum_{l=1}^{d'} x_l^2} \sqrt{\sum_{lp=1}^{d'K} f_{lp}^2} = \norm{x}^2\norm{f}^2$}
    &\nonumber\le \sqrt{\substack{\displaystyle \Lambda^2\norm{f}^2 +
        2\Lambda \norm{x}\norm{f}\sqrt{\sum_{p=1}^K (s_p - 1_{[y=p]})^2} \\
        \displaystyle\quad + ( \sum_{l=1}^{d'} x_l^2 ) ( \sum_{p=1}^K (s_p - 1_{[y=p]})^2 )
    }}
    \intertext{since $\max_{s_1,\dots,s_K}\big\{(s_p-1)^2 + \sum_{q=1,q\neq p}^K s_q^2 \mid \sum_{k=1}^K s_k = 1 \land \forall k \colon s_k \ge 0\big\} = 2$ with $s_q=1 \land  s_p=0~\bigwedge_{k=1, k \neq q}^K s_k = 0$ where $q\neq p$}
    &\le \sqrt{ \Lambda^2\norm{f}^2 +
        2\sqrt{2}\Lambda \norm{x}\norm{f} + 2\norm{x}^2
    }
    = \Lambda\norm{f} + \sqrt{2}\norm{x}
\end{align}

Thus, with $\norm{x} \le c, \norm{f} \le R$ we conclude that
\begin{align}
    &\nonumber\phantom{\le} \sup_{z\in D, f,f'} \frac{\norm{\mathcal{J}(f,z) - \mathcal{J}(f',z)}}{\norm{f - f'}} \\
    &\nonumber\le \sup_{z\in D, f} \norm{\nabla_f\mathcal{J}(f,z)} \\
    &\le \Lambda R + \sqrt{2}c
    = L
\end{align}
\end{proof}

\subsubsection{Proof of \Cref{thm:smooth_softmax} (Smoothness of \softmaxslp)}
\label{app:smooth_softmax}

We state the full version of \Cref{thm:smooth_softmax}:
\begin{reptheorem}{thm:smooth_softmax}
    Let $\mathcal{J}(f,D) \coloneqq \frac\Lambda2 \sum_{k=1}^K \inner{f_k}{f_k} + \frac1n\sum_{(x,y)\in D} \mathcal{L}_{\mathrm{CE}}(y, \inner{f}{x})$ denote the objective function with the cross-entropy loss $\mathcal{L}_{\mathrm{CE}}(y, z) \coloneqq -\sum_{k=1}^K y_k\log\frac{\exp z_k}{\sum_{j=1}^K \exp z_j}$ and parameters $f\in\mathbb{R}^{d+1,K}$, dataset $D$ where $(x,y)\in D$ with data points $x\in\mathbb{R}^{d+1}$ structured as $\begin{bmatrix} 1 & x_1 & \dots & x_d \end{bmatrix}$ and labels $y\in\set{0, 1}^K$, number of classes $K$, and regularization parameter $\Lambda$. $\mathcal{J}$ is $\beta$-smooth with $\beta=\sqrt{(d+1)K \Lambda^2 + 0.5(\Lambda + c^2)^2}$ where $\norm{x} \le c$.

\end{reptheorem}

\begin{proof}
In the following, we abbreviate $d' \coloneqq d+1$, flatten $f \in \mathbb{R}^{d'K}$ and notate $z \coloneqq (x,y)$.

$\beta$-Smoothness is defined as
\[
    \textstyle\sup_{z\in D, f,f'}\frac{\norm{\nabla_f\mathcal{J}(f,z) - \nabla_{f'}\mathcal{J}(f',z)}}{\norm{f - f'}} \le \beta.
\]

We first (1) show
\begin{align}
    &\nonumber\phantom{\le} \sup_{z\in D, f,f'}\frac{\norm{\nabla_f\mathcal{J}(f,z) - \nabla_{f'}\mathcal{J}(f',z)}}{\norm{f - f'}} \\
    &\le \sup_{z\in D, f}\norm{\mathbf{H}_{f}(\mathcal{J}(f,z))}
\end{align}
using the mean value theorem and subsequently (2) bound $\sup_{z\in D, f}\norm{\mathbf{H}_{f}(\mathcal{J}(f,z))} \le \beta$.

(1) Recall that the multivariate mean value theorem states that for some function $g\colon G \mapsto \mathbb{R}$ on an open subset $G \in \mathbb{R}^n$, some $x,y \in G$ and some $c \in [0,1]$, we have
\begin{align}
    g(y) - g(x) = \inner{\nabla g((1-c)x + cy)}{y-x}.
\end{align}

In our case, we write
\begin{align}
    &\nonumber\textstyle\phantom{=}~\sup_{z\in D, f,f'}\frac{\norm{\nabla_f\mathcal{J}(f,z) - \nabla_{f'}\mathcal{J}(f',z)}}{\norm{f-f'}}\\
    &\nonumber\textstyle= \sup_{z\in D, f,f'}\frac{\sqrt{\sum_{i=0}^{d'K}(\nabla_{f_i}\mathcal{J}(f,z) - \nabla_{f'_i}\mathcal{J}(f',z))^2}}{\norm{f-f'}}
    \intertext{by the multivariate mean value theorem for some $c \in [0,1]$ and $g_i(f,z) \coloneqq \nabla_{f_i} \mathcal{J}(f,z)$}
    &\nonumber\textstyle= \sup_{z\in D, f,f'}\frac{\sqrt{\sum_{i=0}^{d'K}\inner{\nabla g_i((1-c)f' - cf,z)}{f-f'}^2}}{\norm{f-f'}}
    \intertext{for $f'' \coloneqq (1-c)f' - cf$ and by the Cauchy-Schwarz inequality $\abs{\inner{\nabla g_i(f'', z)}{f-f'}}^2 \le \norm{\nabla g_i(f'', z)}^2 \cdot \norm{f-f'}^2$}
    &\nonumber\textstyle\le \sup_{z\in D, f''}\sqrt{\sum_{i=0}^{d'K}\sum_{j=0}^{d'K}(\nabla^2_{f''_i,f''_j} \mathcal{J}(f'',z))^2}\\
    &\textstyle= \sup_{z\in D, f}\norm{\mathbf{H}_{f}(\mathcal{J}(f,z))}.
\end{align}

(2) We know that with $1 \le l \le d', 1 \le p \le K$ the first-order partial derivative of $\mathcal{J}$ is $\frac{\partial}{\partial f_{lp}}\mathcal{J}(f,(x,y)) = \Lambda f_{lp} + x_l\cdot(s_p - 1_{[y=p]})$ with $s_p \coloneqq \frac{\exp \inner{f_p}{x}}{\sum_{i=1}^K \exp \inner{f_i}{x}}$.

With $1 \le j \le d', 1 \le q \le K$ we know that the second-order partial derivative of $\mathcal{J}$ is $\frac{\partial^2}{\partial f_{lp} \partial f_{jq}}\mathcal{J}(f,(x,y)) = 1_{[lp=jq]}\cdot\Lambda + x_l\cdot x_j \cdot s_p(1_{[p=q]} - s_q)$.
Thus, we have
\begin{align}
    &\nonumber\phantom{=~}\norm{\mathbf{H}_f(\mathcal{J}(f,z))} \\
    &\nonumber= \sqrt{\sum_{lp=1}^{d'K}\sum_{jq=1}^{d'K}\left(
            1_{[lp=jq]}\cdot\Lambda + x_l x_j s_p(1_{[p=q]} - s_q)
    \right)^2}\\
    &\nonumber= \sqrt{\substack{\displaystyle \sum_{lp=1}^{d'K}\biggl(
            (\Lambda + x_l^2 s_p(1 - s_p))^2 + \sum_{\substack{jq=1\\ j \neq l}}^{d'K} x_l^2 x_j^2 s_p^2(1 - s_p)^2 \\
            \displaystyle\qquad\qquad\qquad\qquad\quad + \sum_{\substack{jq=1\\ j \neq l\\ q \neq p}}^{d'K} x_l^2 x_j^2 s_p^2 s_q^2
    \biggr)}}\\
    &\nonumber= \sqrt{\substack{\displaystyle\sum_{lp=1}^{d'K}\biggl(
            (\Lambda + x_l^2 s_p(1 - s_p))^2\qquad\qquad\qquad \\
            \displaystyle\qquad + x_l^2 s_p^2 \sum_{\substack{j=1\\ j \neq l}}^{d'} \Bigl( x_j^2 (1 - s_p)^2 + x_j^2 \sum_{\substack{q=1\\ q \neq p}}^{K} s_q^2 \Bigr)
    \biggr)}}
    \intertext{since we have $\max_{s_1,\dots,s_K}\big\{ \sum_{q=1,q\neq p}^K s_q^2 \mid \sum_{q=1,q\neq p}^K s_q = 1 - s_p \land \forall i \colon s_i \geq 0\big\} = (1 - s_p)^2$ due to the maximal L2-distance given a bounded L1-distance is the maximal L2-distance in one dimension, we conclude}
    &\nonumber= \sqrt{\substack{\displaystyle\sum_{lp=1}^{d'K}\biggl(
            \Lambda^2 + 2\Lambda x_l^2 s_p(1-s_p) + x_l^4 s_p^2(1 - s_p)^2 \\
            \displaystyle + 2x_l^2 s_p^2(1 - s_p)^2 \sum_{\substack{j=1\\ j \neq l}}^{d'}  x_j^2
    \biggr)}}\\
    &\nonumber\leq \sqrt{\substack{\displaystyle d'K \Lambda^2 + \sum_{lp=1}^{d'K}
            x_l^2 s_p(1-s_p) \Bigl(2\Lambda +  2 x_l^2 s_p(1 - s_p) \\
            \displaystyle\qquad\qquad\qquad\qquad\qquad\qquad + 2s_p (1-s_p) \sum_{\substack{j=1\\ j \neq l}}^{d'} x_j^2 \Bigr)}}\\
    &\nonumber= \sqrt{d'K \Lambda^2 + 2\sum_{lp=1}^{d'K}
            x_l^2 s_p(1-s_p) \Bigl(\Lambda + s_p (1-s_p) \norm{x}^2 \Bigr)}\\
    &\nonumber\le \sqrt{d'K \Lambda^2 + 2\norm{x}^2 \sum_{l=1}^{d'} x_l^2 \sum_{p=1}^{K}
            s_p(1-s_p) (\Lambda\norm{x}^{-2} + s_p)
    }
    \intertext{following \Cref{lem:max_smoothness} (presented and shown below) we simplify with $C \coloneqq \Lambda\norm{x}^{-2}$: $\sum_{p=1}^{K} s_p(1-s_p) (C + s_p) \le 0.25(C + 1)^2$}
    &\nonumber\le \sqrt{d'K \Lambda^2 + 0.5\norm{x}^4
            (\Lambda\norm{x}^{-2} + 1)^2
    } \\
    &= \sqrt{d'K \Lambda^2 +
            0.5(\Lambda + \norm{x}^{2})^2
    }.
\end{align}

Thus, with $\norm{x} \le c$ we conclude that
\begin{align}
    &\nonumber\phantom{\le~}
        \sup_{z\in D, f,f'}\frac{\norm{\nabla_f\mathcal{J}(f,z) - \nabla_{f'}\mathcal{J}(f',z)}}{\norm{f - f'}} \\
    &\nonumber\le
        \sup_{z\in D, f}\norm{\mathbf{H}_{f}(\mathcal{J}(f,z))} \\
    &\le
        \sqrt{d'K \Lambda^2 + 0.5(\Lambda + c^2)^2} = \beta.
\end{align}
\end{proof}

\begin{lemma}
\label{lem:max_smoothness}
    Let $\set{s_p}_{p=1}^K$ denote probabilities such that $\sum_{p=1}^K s_p = 1$, and $C \in \mathbb{R}_+$ a constant, then we have
\begin{align}
    \nonumber\max_{\{s_p\}_{p=1}^K} &\Big\{ \sum_{p=1}^K s_p(1 - s_p)(C + s_p) \mid \sum_{p=1}^K s_p = 1 \land \forall p \colon s_p \ge 0 \Big\}\\
    &\le 0.25(C+1)^2
\end{align}
with $\forall{p \in \cup_{i=1}^k P_i, p' \in \cup_{i=k+1}^K P_i, P \in \operatorname{Sym}(K)}\colon (s_{p} = \frac1k \land s_{p'} = 0)$, i.e. for some arbitrary but fixed dimensions $k: 1 \le k \le K$, the solution has $k$-times $s_p=\frac1k$ and $(K-k)$-times $s_p=0$.

\end{lemma}

\begin{proof}
We show this Lemma as follows: First, we use the Karush–Kuhn–Tucker (KKT) conditions to find the $s_p$'s which maximize the maximization term. Thereby, we obtain a set of four solution candidates where we encode all $s_p$'s in closed form and introduce two new variables $k,j$ which serve as a solution counter.
Second, we insert the solution candidates into the maximization term and show that the result is always bounded by $0.25(C+1)^2$ by calculating the optimal front across all possible values of the solution counters $k,j$.

Let $f(s) \coloneqq \sum_{p=1}^K s_p(1-s_p)(C+s_p)$ denote the function to maximize, $h(s) \coloneqq \sum_{p=1}^K s_p - 1$ the equality constraint, and $\forall p \colon g_p(s) \coloneqq -s_p$ the inequality constraints. To find the constrained maximum, we maximize the Lagrangian function $\mathcal{L}_{\mathrm{agrange}}(s) =  f(s) + \mu_pg_p(s) + \lambda h(s)$ with $\mu_p, \lambda$ as slack variables.
This suffices since $s_p$ does not have unbounded border cases: the only valid configuration of all $s_p$'s is on a hyperplane ($\sum_p s_p = 1$) bounded in all dimensions ($s_p \ge 0$). Using the slack variable $\mu_p$, we already cover whether its corresponding $s_p$ is on the border ($\mu_p > 0$) or not ($\mu = 0$). 
Following the KKT conditions, the following conditions have to hold for the maximum:
\begin{enumerate}[label=(\arabic*)]
    \item Stationarity: $\forall p \colon \nabla_{s_p}\mathcal{L}_{\mathrm{agrange}}(s) = C + 2s_p - 2Cs_p - 3s_p^2 + \mu_p - \lambda = 0$
    \item Primal feasibility: $\forall p \colon h(s) = 0$ and $g_p(s) \le 0$
    \item Dual feasibility: $\forall p \colon \mu_p \ge 0$
    \item Complementary slackness: $\forall p \colon \mu_p g_p(s) = 0$
\end{enumerate}

Informally, it suffices for the solution of the KKT conditions to analyze the cases where $\forall{p,1 \le p\le k}\colon s_p > 0$ for all fixed number of dimensions $k: 1 \le k \le K$ since if $s_p = 0$ then we have already proved the same result for one less dimension.

Formally and without loss of generality\footnote{The same argumentation holds for situations where the dimensions are permuted.}, we show for all fixed numbers of dimensions $k: 1 \le k \le K$ that for the solution of the KKT conditions it suffices to analyze the cases where $\forall{p,1 \le p\le k}\colon s_p > 0$.
For the induction base case ($k=1$ dimensional), we have $s_1>0$ and thus by condition (4) $\mu_1 = 0$. If and only if $s_1=1$, we satisfy conditions (2) and (1) with $\lambda = -C-1$. With $s_1 = 0$, we would not be able to satisfy the equality constraint of condition (2), i.e., $s_1=1$.\\
For the $k\mapsto k+1$ induction case, we know that $\forall{p,1 \le p\le k}\colon s_p > 0$. If $s_{p+1} > 0$, by the induction hypothesis we know that $\forall{p,1 \le p\le k+1}\colon s_p = 0$. If $s_{p+1} = 0$ then by conditions (3) and (4) we have $\mu_{p+1} > 0$ and thus by condition (1), $\mu_{p+1} = \lambda - C$. Inserting $s_{p+1}=0, \mu_{p+1}=\lambda-C$ into conditions (1) to (4), we obtain the same set of equations and inequalities as for the $k$-dimensional case which already holds by the induction hypothesis.

We solve the KKT conditions (1) to (4) as follows:
First, we solve the system of equations of condition (1) for $s_p$ via the quadratic formula:
\begin{align}\label{eq:sp_pm}
    \textstyle s_p^{\pm} &= \frac{-(2-2C) \pm \sqrt{(2-2C)^2 - 4(-3)(C-\lambda)}}{2(-3)} \\
    &= \sfrac{1}{3}\cdot\left(\pm\sqrt{C^2+C-3\lambda+1} - C + 1\right).
\end{align}
Second, we plug $s_p^{\pm}$ into the equality constraint, $h(s) = 0$, of condition (2) and solve for $\lambda$ which gives us for some solution counter $j\in\mathbb{N}, 0 \le j \le k$ with $2j \neq k$:
\begin{align}
    &\nonumber\phantom{\iff~} \textstyle h(s^{\pm})=0\\
    &\nonumber\iff\textstyle (\sum_{i=1}^j s_i^+) + (\sum_{i=j+1}^k s_i^-) = 1 \\
    &\nonumber\iff\textstyle j \left(\sqrt{C^2+C-3\lambda+1} - C + 1\right) \\
    &\nonumber\phantom{\iff} + (k-j) \left(-\sqrt{C^2+C-3\lambda+1} - C + 1\right) = 3 \\
    &\nonumber\iff\textstyle (2j-k) \sqrt{C^2+C-3\lambda+1} = Ck - k + 3 \\
    &\nonumber\phantom{\Leftarrow}\Rightarrow\ \ \textstyle C^2 + C - 3\lambda + 1 = \frac{(Ck - k + 3)^2}{(2j - k)^2} \\
    &\iff\textstyle \lambda = \frac{(2j - k)^2 (C^2 + C + 1) - (Ck - k + 3)^2}{3(2j - k)^2}.
\end{align}

The solution counter $j$ quantifies how often we plug the `positive' variant of $s_p^{\pm}$ into $h(s^{\pm})$:
\begin{align}
    \textstyle s^{\pm} \coloneqq \begin{bmatrix}
        s_1^{+} & \dots & s_j^{+} & s_{j+1}^{-} & \dots & s_k^{-}
    \end{bmatrix}
\end{align}

or any permutation of the dimensions of $s^{\pm}$.

Note that at $2j = k$, we have a special case and by the equality constraint, $h(s) = 0$, of condition (2)
\begin{align}
    &\nonumber\phantom{\iff~}\textstyle h(s^{\pm}) = 0~\land~2j=k \\
    &\nonumber\textstyle\iff (\sum_{p=1}^{\frac{k}{2}} s_p^+) + (\sum_{p=\frac{k}{2}+1}^k s_p^-) = 1 \\
    &\nonumber\textstyle\iff k(1-C) =3 \\
    &\textstyle\iff C = \frac{k-3}{k}.
\end{align}
Thus, at $2j = k, C= \frac{k-3}{k}$ we simplify the solution in \Cref{eq:sp_pm} to
\begin{align}
    \nonumber\textstyle s_p^{\pm,~C=\sfrac{k-3}{k}} &\textstyle= \sfrac13 \cdot (\pm \underbrace{\textstyle\sqrt{\frac{(k-3)^2}{k^2} + \frac{k-3}{k} - 3\lambda + 1}}_{\eqqcolon~3Q} - \frac{k-3}{k} + 1) \\
    &\textstyle= \pm Q + \frac{1}{k}.
\end{align}
If we now insert $s_p^{\pm,~C=\sfrac{k-3}{k}}$ into $f(\cdot)$ and maximize for all remaining variables, we find the maximum at
\begin{align}
    &\nonumber\textstyle\phantom{=}~\max_{k, j, \lambda} \Big\{ f(s_p^{\pm,~C=\sfrac{k-3}{k}}) \\
    &\nonumber\textstyle\phantom{= \max_{k, j, \lambda} \Big\{} \mid 2j=k \land C= \frac{k-3}{k} \land s_p^{\pm,~C=\sfrac{k-3}{k}} \ge 0 \Big\} \\
    &\nonumber\textstyle\le \max_{k, j, \lambda} \big\{ \sum\limits_{p=1}^{\frac{k}{2}} (\frac{1}{k} + Q)(1 - (\frac{1}{k} + Q))(C + (\frac{1}{k} + Q)) \\
    &\nonumber\textstyle\phantom{\le \max_{k, j, \lambda} \big\{} + \sum\limits_{p=\frac{k}{2}+1}^{k} (\frac{1}{k} - Q)(1 - (\frac{1}{k} - Q))(C + (\frac{1}{k} - Q)) \\
    &\nonumber\textstyle\phantom{\le \max_{k, j, \lambda} \big\{} \mid 2j=k \land C= \frac{k-3}{k} \big\} \\
    &\nonumber\textstyle= \max_{k} \myset{ \frac{k}{2} \frac{1}{k}(1 - \frac{1}{k})(\frac{k-3}{k} + \frac{1}{k}) + \frac{k}{2} \frac{1}{k}(1 - \frac{1}{k})(\frac{k-3}{k} + \frac{1}{k}) } \\
    &\nonumber\textstyle= \max_{k} \myset{ (1 - \frac{1}{k})(\frac{k-3}{k} + \frac{1}{k}) } \\
    &\nonumber\textstyle= \max_{k} \myset{ \frac{k-2}{k} - \frac{k-2}{k^2} } \\
    &\textstyle= \max_{k} \big\{ \underbrace{\textstyle 1 - \frac{3}{k} + \frac{2}{k^2}}_{ \le~0.25(\frac{k-3}{k}+1)^2~=~1 - \frac{3}{k} + \frac{9}{4k^2} } \big\}.
\end{align}
Thus, at $2j=k$, $\mathcal{L}_{\mathrm{agrange}}$ is maximal at $C=\frac{k-3}{k}$ which is always strictly below the maximum we will show in this lemma if $C=\frac{k-3}{k}$. In the following, we continue the proof for $2j\neq k$.

Third, by plugging $\lambda$ into \Cref{eq:sp_pm} which is derived from the system of equations in condition (1) and solving for $s_p$, we obtain the following two solution candidates for $2j \neq k$
\begin{gather}
    \begin{aligned}
    &\nonumber\textstyle\phantom{=~} s_p^{(+,-)} \\
    &\nonumber= \textstyle\frac{1}{3}\big( \pm\sqrt{C^2 + C - 3 \frac{(2j - k)^2 (C^2 + C + 1) - (Ck - k + 3)^2}{3(2j - k)^2} + 1} \\
    &\nonumber\textstyle\phantom{= \frac{1}{3}\big(} - C + 1 \big) \\
    &\nonumber= \textstyle \frac{1}{3} \big(1 - C \pm \frac{Ck - k + 3}{2j - k} \big) \\
    &\nonumber\textstyle= \frac{(2j - k) (1 - C) \pm (Ck - k + 3)}{3(2j - k)} \\
    &\nonumber\textstyle= \frac{-2Cj + Ck + 2j - k \pm(Ck - k + 3)}{6j - 3k}
    \end{aligned}\\
    \textstyle s_p^{(+)} = \frac{- 2(k-j)C + 2(k -j) - 3}{6(k-j)-3k},
    s_p^{(-)} = \frac{-2jC + 2j - 3}{6j - 3k}.
\end{gather}

Observe that if we replace $\Tilde{j} \coloneqq k - j$ in $s_{p}^{(+)}$ we get $s_{p}^{(-)}$ with $\Tilde{j}$ instead of $j$.
To abbreviate, we write
\begin{align}
    \textstyle s_p^{(j')} = \frac{-2j'C + 2j' - 3}{6j' - 3k}
\end{align}
for $j' \in \myset{j, k-j}$.
Because of the similar structure of $s_p^{(j)}$ and $s_p^{(k-j)}$, restricting $j$ by $0 \le 2j < k$ suffices since we would otherwise count the same maximum twice.
With $s_p^{(j')}$ as our solution candidate, the equality constraint, $h(s)= 0$, in condition (2) holds when we have $(k-j)$ times $s_p^{(j)}$ and $j$ times $s_p^{(k-j)}$:
\begin{align}
    \textstyle s^{\mathrm{sol}} \coloneqq \begin{bmatrix}
        s_1^{(k-j)} & \dots & s_j^{(k-j)} & s_{j+1}^{(j)} & \dots & s_k^{(j)}
    \end{bmatrix}
\end{align}
or any permutations of the dimensions of $s^{\mathrm{sol}}$. This goes by construction of $s^{\pm}$ where the solution counter $j$ quantifies how often we plug in $s_p^{(+)}$ into $h(s^{(+,-)})$.

We next compute the second partial derivative test to determine for which parameters the solution candidate $s^{\mathrm{sol}}$ is a local maximum or minimum: We have a maximum if the Hessian of $\mathcal{L}_{\mathrm{agrange}}$ is positive definite and a minimum if the Hessian of $\mathcal{L}_{\mathrm{agrange}}$ is negative definite. In our case, the second partial derivatives of $\mathcal{L}_{\mathrm{agrange}}$ are $\nabla^2_{s_p}\mathcal{L}_{\mathrm{agrange}}(s) = 2-2C-6s$ and $\nabla_{s_p}\nabla_{s_q}\mathcal{L}_{\mathrm{agrange}}(s) = 0$ with $p \neq q$. Thus, we have a diagonal Hessian matrix. Hence, if $2-2C-6s^{\mathrm{sol}} < \boldsymbol{0}$ we have a maximum and if $2-2C-6s^{\mathrm{sol}} > \boldsymbol{0}$ we have a minimum.
Because of the second partial derivative test, we also know that if the Hessian has both positive and negative eigenvalues, then we have a saddle point. This holds in our case when we have both positive and negative values on the diagonals of the Hessian, i.e., for some $p$ we have $2-2C-6s^{\mathrm{sol}}_p < 0$ and for some $q$ we have $2-2C-6s^{\mathrm{sol}}_q > 0$. Furthermore, if we have a zero eigenvalue, this test is indecisive.

We rearrange the maximum condition for any entry of $s^{\mathrm{sol}}$ (here: $s_p^{(j')}$) as follows:
\begin{align}
    &\nonumber\textstyle\phantom{\iff~} 2-2C-6\frac{-2j'C+2j'-3}{6j'-3k} < 0 \\
    &\nonumber\iff\textstyle \begin{cases}
        kC - k + 3 > 0 &\mathrm{if }\textstyle 0\le 2j'< k \\
        kC - k + 3 < 0 &\mathrm{if }\textstyle 2j'> k
    \end{cases} \\
    &\iff\textstyle \begin{cases}
        C > \frac{k-3}{k} &\mathrm{if }\textstyle 0\le 2j'< k \\
        C < \frac{k-3}{k} &\mathrm{if }\textstyle 2j'> k
    \end{cases}.
\end{align}

Similarly, we rearrange the minimum condition, such that
\begin{align}
    &\nonumber\textstyle\phantom{\iff~} 2-2C-6\frac{-2j'C+2j'-3}{6j'-3k} > 0 \\
    &\iff\textstyle \begin{cases}
        C < \frac{k-3}{k} &\mathrm{if }\textstyle 0\le 2j'< k \\
        C > \frac{k-3}{k} &\mathrm{if }\textstyle 2j'> k
    \end{cases}.
\end{align}

Recall that at this point we only consider $2j \neq k$. 
We now distinguish three cases for the second partial derivative test for the vector $s^{\mathrm{sol}}$: $C<\frac{k-3}{k}$, $C>\frac{k-3}{k}$, $C=\frac{k-3}{k}$.

At $C<\frac{k-3}{k}$, we write
\begin{align}
    \textstyle\begin{bmatrix}
        2-2C-6 s_1^{(k-j)} < 0 \\ \dots \\ 2-2C-6 s_j^{(k-j)} < 0 \\ 2-2C-6 s_{j+1}^{(j)} > 0 \\ \dots \\ 2-2C-6 s_k^{(j)} > 0
    \end{bmatrix}
\end{align}
and at $C>\frac{k-3}{k}$, we write similarly
\begin{align}
    \textstyle\begin{bmatrix}
        2-2C-6 s_1^{(k-j)} > 0 \\ \dots \\ 2-2C-6 s_j^{(k-j)} > 0 \\ 2-2C-6 s_{j+1}^{(j)} < 0 \\ \dots \\ 2-2C-6 s_k^{(j)} < 0
    \end{bmatrix}.
\end{align}

Recall the saddle point criteria as $\exists_p \exists_q~2-2C-6s^{\mathrm{sol}}_p < 0 \land 2-2C-6s^{\mathrm{sol}}_q > 0$ and the maximum criteria as $2-2C-6s^{\mathrm{sol}} < \boldsymbol{0}$.
By the above test criteria, for $C \neq \frac{k-3}{k}$, we have a saddle point for all $j \in [1,k-1]$ as well as a maximum for $j=k \land C<\frac{k-3}{k}$ and for $j=0 \land C>\frac{k-3}{k}$ at
\begin{align}
    \nonumber\textstyle s^{\mathrm{max}} &\coloneqq \begin{bmatrix} s_1^{(k-k)} & \dots & s_k^{(k-k)} \end{bmatrix} \\
    &\nonumber= \begin{bmatrix} s_1^{(0)} & \dots & s_k^{(0)} \end{bmatrix} \\
    &= \begin{bmatrix} \sfrac{1}{k} & \dots & \sfrac{1}{k} \end{bmatrix} \land C \neq \frac{k-3}{k}
\end{align}
since only at $j \in \myset{0, k}$ do we have the case that either $s_p^{(j)}$ or $s_p^{(k-j)}$ is present in the solution $s^{\mathrm{sol}}$.

At $C=\frac{k-3}{k}$, we have for any entry of $s^{\mathrm{sol}}$ (here: $s_p^{(j')}$)
\begin{align}
    s_p^{(j', C=\sfrac{k-3}{k})} = \frac{-2(k-3)\sfrac{j'}{k} + 2j' - 3}{6j' - 3k} = \frac{6\sfrac{j'}{k} - 3}{6j' - 3k} = \frac{1}{k}.
\end{align}
Thus, although the second partial derivative test is indecisive since $2-2C-6\sfrac{1}{k} = 0$, we have at $C=\frac{k-3}{k}$ always the same solution as in $s^{\mathrm{max}}$. This renders $s^{\mathrm{max}}$ for all $C$ as the maximal solution.

Next, we plug the solution $s^{\mathrm{max}}$ into $f(s)$ and calculate the optimal front with the inequality constraint, $g_p(s) \le 0$, of condition (2) and across all number of dimensions $k$ and range of the solution counter $j \in \myset{0, k}$:
\begin{align}
    &\nonumber\max_{k,j} \Big\{\, \textstyle f(s^{\mathrm{max}})  \mid s_p^{(j)} \ge 0 \land s_p^{(k-j)} \ge 0 \land j \in \myset{0, k} \,\Big\} \\
    =~&\nonumber\max_{k} \Big\{\, \textstyle \sum_{p=1}^k s_p^{(0)} (1 - s_p^{(0)}) (C + s_p^{(0)}) \mid s_p^{(0)} \ge 0 \,\Big\}\\
    =~&\nonumber\max_{k} \Big\{\, \textstyle \sum_{p=1}^k \frac{1}{k} (1 - \frac{1}{k}) (C + \frac{1}{k}) \mid \frac{1}{k} \ge 0 \,\Big\}\\
    =~&\nonumber\max_{k} \Big\{\, \textstyle C + \frac{1 - C}{k} - \frac{1}{k^2} \,\Big\}
    \intertext{\big(for $k=\frac{2}{1-C}$ the term $C + \frac{1 - C}{k} - \frac{1}{k^2}$ is maximal for which we need the derivative to be zero: $\frac{d}{dk}(C + \frac{1 - C}{k} - \frac{1}{k^2}) = \frac{C-1}{k^2} + \frac{2}{k^3} = 0$\big)}
    =~&\nonumber\textstyle C + \frac12(1-C)^2 - \frac14(1-C)^2\\
    =~&\textstyle \frac{C^2}{4} + \frac{C}{2} + \frac14 = 0.25(C+1)^2
\end{align}

Thus, we conclude that $f(s^{\mathrm{max}})$ is equal to or below the convex hull $0.25(C+1)^2$ for any solution counter $j$ and any number of classes $k$.
\end{proof}

Note: In this proof, we assumed $k \in \mathbb{R}_+$, however, we can restrict the number of classes $k$ even further: $k \in \mathbb{N}$ and $k \le K$. Yet, this restriction does not have much impact on the bound on $f$ for a reasonable $C, K$: Now, we only have $K$ possible maxima ($s_p^{\mathrm{max}}=\myset{1,\sfrac12,\dots,\sfrac1K}$) where for a given $C$ only one of these maxima are dominant. This also means that our $0.25(C+1)^2$-bound is a convex hull and only matches the maxima in a few selected points. However, already for little $K$ does the maximum come considerably close to the hull as shown in \Cref{fig:viz_softmax_smooth}.

\begin{figure}[!ht]
    \centering
    \centerline{\resizebox{\columnwidth}{!}{\input{plots/viz_softmax_smooth.pgf}\unskip}}
    \caption{
        Precise maximum of $f(s^{\mathrm{max}})$ per constant $C$ and restricted, discretized number of classes $k \le K, k \in \mathbb{N}$ versus convex hull of the maximum of $f(s^{\mathrm{max}})$ across all number of classes $k \in \mathbb{R}_+$.
    }
    \label{fig:viz_softmax_smooth}
\end{figure}

\end{document}